\documentclass[pdflatex,sn-mathphys]{sn-jnl}

\usepackage[utf8]{inputenc}
\usepackage{amsmath,amssymb}
\usepackage{graphicx}
\usepackage{subfig}
\usepackage{caption}
\usepackage{url}
\usepackage[numbers]{natbib} 
\usepackage{placeins}
\usepackage{xcolor}
\usepackage{siunitx}
\usepackage{url}

\usepackage{xspace}

\usepackage{enumitem}

\usepackage{booktabs} 
\usepackage{multirow}
\usepackage{multicol}
\usepackage{makecell}

\usepackage{etoolbox}

\usepackage{setspace}

\definecolor{wisconsin-red}{rgb}{0.6,0,0}
\definecolor{darkgreen}{rgb}{0.2,0.6,0.2}
\definecolor{maroon}{rgb}{0.5, 0.0, 0.0}
\definecolor{violet}{rgb}{0.75, 0.0, 1.0}
\definecolor{lightgray}{gray}{0.9}
\definecolor{navyblue}{rgb}{0.0, 0.0, 0.5}
\definecolor{darkmidnightblue}{rgb}{0.0, 0.2, 0.4}
\definecolor{midnightblue}{rgb}{0.0,0.4,0.85}
\definecolor{Gray}{gray}{0.75}
\definecolor{darkgreen}{rgb}{0,0.5,0}
\definecolor{apricot}{rgb}{0.98, 0.81, 0.69}

\newcolumntype{C}[1]{>{\centering\arraybackslash}p{#1}}
\newcolumntype{P}[1]{>{\raggedright\arraybackslash}p{#1}}

\newcommand{\bertLarge}{\ensuremath{\text{BERT}_\text{LARGE}}}
\newcommand{\robertaLarge}{\ensuremath{\text{RoBERTa}_\text{LARGE}}}
\newcommand{\debertaLarge}{\ensuremath{\text{DeBERTa}_\text{LARGE}}}

\newcommand{\instanceVocabulary}{t}
\newcommand{\mask}{\underline{\hspace{0.75cm}}}
\newcommand{\segsep}{\ $\Vert$\ }
\newcommand{\setModel}{\mathcal{M}}
\newcommand{\instanceModel}{m}
\newcommand{\setLabel}{\mathcal{L}}
\newcommand{\instanceLabel}{\ell}

\newcommand{\textA}{\ensuremath{u}}
\newcommand{\textB}{\ensuremath{v}}
\newcommand{\petVerbalizer}{v}
\newcommand{\petEx}{x}
\newcommand{\petPattern}{p}
\newcommand{\petScore}{s}
\newcommand{\petWeight}{w}

\newcommand{\setTrainData}{\mathcal{T}}

\newcommand{\setFitNumIterations}{R}

\newcommand{\setfitMpnet}{$\textsc{SetFit}_\textsc{MPNet}$}
\newcommand{\fineTuneSetFit}{$\textsc{FineTune}_\textsc{SetFit}$}
\newcommand{\fineTune}{$\textsc{FineTune}$}

\newfloat{lstfloat}{htbp}{lop}
\floatname{lstfloat}{Listing}

\begin{document}

\title[Few-shot learning for sentence pair classification]{Few-shot learning for sentence pair classification and its applications in software engineering}

\author[1]{\fnm{Robert Kraig} \sur{Helmeczi}}
\author*[1]{\fnm{Mucahit} \sur{Cevik}}\email{mcevik@torontomu.ca}
\author[3]{\fnm{Savas} \sur{Yildirim}}

\affil[1]{
\orgname{Toronto Metropolitan University}, \orgaddress{\street{44 Gerrard St E}, \city{Toronto}, \postcode{M5B 1G3}, \state{Ontario}, \country{Canada}}}

\abstract{
Few-shot learning---the ability to train models with access to limited data---has become increasingly popular in the natural language processing (NLP) domain, as large language models such as GPT and T0 have been empirically shown to achieve high performance in numerous tasks with access to just a handful of labeled examples.
Smaller language models such as BERT and its variants have also been shown to achieve strong performance with just a handful of labeled examples when combined with few-shot learning algorithms like pattern-exploiting training (PET) and SetFit.
The focus of this work is to investigate the performance of alternative few-shot learning approaches with BERT-based models.
Specifically, vanilla fine-tuning, PET and SetFit are compared for numerous BERT-based checkpoints over an array of training set sizes.
To facilitate this investigation, applications of few-shot learning are considered in software engineering.
For each task, high-performance techniques and their associated model checkpoints are identified through detailed empirical analysis.
Our results establish PET as a strong few-shot learning approach, and our analysis shows that with just a few hundred labeled examples it can achieve performance near that of fine-tuning on full-sized data sets.  
}


\maketitle

\section{Introduction}

With the recent development of powerful pre-trained language models such as BERT \citep{devlin2019bert} and GPT-3 \citep{brown2020gpt3}, many studies have started to investigate the impact of few-shot learning, the practice of training high-quality models using just a handful of labeled training instances.
Few-shot learning represents a significant source of savings, both in terms of time and money, for a variety of classification tasks.
For instance, in software engineering, detecting bug dependencies and duplicates can be an arduous task for bug triagers.
In general, for a repository with $n$ bugs, detecting dependencies and duplicates requires comparing all pairs of bugs, a task which has $\mathcal{O}(n^2)$ complexity.
For large $n$, this can become intractable for even a computer to solve, but even for relatively small values of $n$, manual comparisons become extremely time consuming.
Training a language model to detect these relationships can be an effective approach for automating this procedure, but this generally requires manually labeling tens of thousands of bugs as duplicates or dependent.
Labeling such a large number of bug reports is costly, requiring enormous amounts of time and money to be dedicated to this task.
For many development projects, even if these costs could be absorbed, there may not be enough bug reports to label.
In such problems, training a language model without labeling a substantial amount of training data offers a significant advantage.

With this description of few-shot learning, one quickly notes that it is \textit{always} preferable to be able to train high-quality models with just a handful of labeled training data.
However, few-shot learning might come with a decrease in model accuracy, and the tolerability of such decreases changes with both the task and the overseeing entity.
For example, in the automatic essay scoring domain, contract graders are expected to grade with an accuracy between 80\% and 95\%~\citep{blees.ai}.
Accordingly, it is important to quantify the ability of few-shot learning methods on a variety of tasks to inform the decisions of industry professionals seeking to apply them in their field. 

This research focuses on the applications of few-shot learning algorithms in important practical problems in the domain of software engineering.
In particular, we focus on pattern-exploiting training (PET) \citep{schick2021exploiting} and SetFit \citep{tunstall2022efficient} as few-shot learning methods, and we compare their performance with vanilla fine-tuning.
Much of the literature on few-shot learning generally involves comparing and contrasting various few-shot approaches using different Transformer model checkpoints, or by focusing on just a single checkpoint \citep{tunstall2022efficient, tam2021adapet, mahabadi2022perfect}.
This is generally done with the intention of emulating a particular few-shot learning setting, where validation data is not available to evaluate an approach's performance for each checkpoint.
Instead, model checkpoints are often chosen based on model performance on full-sized datasets \citep{schick2021size}, but this approach might not always be indicative of few-shot learning performance.
Accordingly, we conduct an empirical analysis with different few-shot learning approaches on a variety of model checkpoints to better inform the selection of the ideal checkpoint.

Another limitation in the current literature is on dataset size.
\citet{alex2021raft} note that, in a few-shot setting, it is generally not desirable to label more than 50 examples for training data, and many studies on few-shot learning consider similarly-small datasets \citep{schick2021exploiting, mahabadi2022perfect, tam2021adapet}.
While labeling more examples may not be preferable, studies that consider only datasets of this size create a huge disparity in the literature between the few-shot learning domain and language modeling on full-size datasets, which often contain thousands or tens-of-thousands of labeled examples.
Accordingly, we investigate the performance of few-shot learning approaches beyond this 50-labeled example limit, providing a thorough analysis of each approach over a range of training set sizes.
The results provide a comprehensive overview of the leading checkpoints and approaches across various small dataset sizes.

Our contributions to text modeling in software engineering can be summarized as follows:
\begin{itemize}
    \item Four text classification tasks are introduced as a test suite to evaluate few-shot learning approaches in the software engineering domain.
          The tasks considered include duplicate detection in bug reports from Bugzilla\footnote{\url{https://bugzilla.mozilla.org}} and questions from Stack Overflow,\footnote{\url{https://stackoverflow.com}} bug dependency detection in bugs from Bugzilla, and conflict detection in software requirement specification (SRS) documents.
          Each of these tasks is formulated as a sentence pair classification problem.
          
    \item PET is adapted to each of the considered software engineering applications and is compared with SetFit and fine-tuning to evaluate the performance of few-shot learning in this domain.
          The ensuing analysis reveals the overall viability of few-shot learning in software engineering applications.
    
    \item The few-shot learning approaches are compared using three popular pre-trained language model checkpoints---\bertLarge{}, \robertaLarge{}, and \debertaLarge{}---as their underlying language model.
         The numerical analysis reveals that fine-tuning with \bertLarge{} is often the best-performing approach when compared to all other checkpoints and approaches in a few-shot setting.
          The ensuing discussion identifies differences in the training procedures for each model checkpoint which may have led to these observations.
          
    \item The performance of each few-shot learning approach is evaluated over a range of dataset sizes.
         The empirical results identify the strengths of each model checkpoint and few-shot approach with respect to training set size, allowing for more informed decision-making when selecting a few-shot learning approach and the language model.
\end{itemize}

The rest of this paper is organized as follows. Section~\ref{sec:lit_review} reviews the literature on the tasks considered, and then discusses the few-shot learning methods that we use to solve them.
Section~\ref{sec:methodology} details the tasks to be solved, the datasets used to represent them, and how we adapt each few-shot learning approach to these tasks.
Section~\ref{sec:results} summarizes the results of the numerical experiments for each of the datasets.
Section~\ref{sec:conclusion} summarizes our findings, identifies limitations and areas of future research, and provides a series of recommendations for applying few-shot learning in a practical setting based on the observations in this work.

\section{Literature Review}\label{sec:lit_review}
In this section, we first briefly review the relevant literature on the software engineering classification tasks considered in this study. Then, we discuss the recent advances in few-shot learning methodologies.
Table~\ref{tab:sentence-pair-lit-review} provides an overview of the most relevant studies for the software engineering tasks considered in our work.

\begin{table}[htb]
    \centering
    \caption{Summary of recent relevant literature on sentence pair classification problems in software engineering.}
    \renewcommand{\arraystretch}{1.5}
    \label{tab:sentence-pair-lit-review}
    \resizebox{0.9\textwidth}{!}{\begin{tabular}{lllcc}
    \toprule 
      Study & Approach & Features & Few-shot? & Task \\
    \midrule
       \citet{aggarwal2017detecting} & Decision-making & C\&T Sim. & No & Bugzilla Duplicate \\
       \citet{chauhan2023denature} & Ranking & TF-IDF vector & No & Bugzilla Duplicate \\
       \citet{kim2022predicting} & Binary Classification & Topic Modeling & No & Bugzilla Duplicate \\[.2cm]
       \midrule
       \citet{wang2020duplicate} & Decision-making & Deep Learning & No & SO Duplicate \\[.2cm]
       \midrule
       \citet{guo2021automatically} & Decision-making & Semantic meta-model & No & SRS \\
       \citet{malik2023transfer} & Decision-making & BERT embeddings & No & SRS\\[.cm]
       \midrule
       Our study & Decision-making & BERT embeddings & Yes & All Tasks\\[.2cm]
    \bottomrule
    \end{tabular}}
    \begin{tablenotes}
    \centering
    \footnotesize
    \item SO: Stack Overflow, SRS: Software Requirement Specification document conflict detection, All Tasks: Entailment + SRS + Bugzilla Duplicate + SO Duplicate, C\&T Sim.: Contextual and Textual Similarity
    \end{tablenotes}
\end{table}

\subsection{Detecting duplicate bug reports}

In the context of bug duplicate detection, \citet{lin2016enhancements} group solution methods into three main categories.
Firstly are ranking approaches which, when given a bug report, return a ranked list of similar matches.
Secondly are binary classification approaches, which train a binary classifier to determine if a bug report is a duplicate or not.
In their work, \citet{lin2016enhancements} found that such classifiers did not offer substantial benefits to triagers as the state-of-the-art still misclassified the majority of duplicates.
Thirdly are decision-making approaches, which take pairs of bug reports and determine if the pair are duplicates.
\citet{chauhan2023denature} provide a detailed literature review of the recent literature on duplicate detection in bug reports.
Accordingly, we refer the reader to their work for an extensive overview on the field.
We summarize here some of the more recent approaches used for duplicate detection.
While in the recent literature, \citet{zhang2022repairing} propose using few-shot learning for bug repairing and \citet{li2022towards} suggest few-shot learning may be helpful for bug entity extraction and relationship modeling, our review of the literature did not identify any research into few-shot learning for bug duplicate detection.
Therefore, we provide a review some of the relevant literature on duplicate detection in a data rich environment.

\citet{chauhan2023denature} use a ranking approach to identify duplicate bug reports.
They use cosine similarity to identify bugs which are similar to one another, choosing a threshold of 0.6, above which bugs are considered to be duplicates.
For vectorization, they use vector space modeling where the feature vector is represented as a TF-IDF vector, achieving prediction accuracy of 88.8\%.
We note that a natural extension of their work would be to apply SBERT~\citep{reimers2019sentence} for generating strong sentence vectors.
\citet{kim2022predicting} employ BERT to create a binary classification model for duplicate bug detection.
They propose using topic modeling for feature extraction, and then using the features for each state of a topic to train a BERT model for determining if a bug report is a duplicate or not.
Their results show substantial improvements over the more common approaches in literature (e.g., CNN, LSTM, RandomForest). 
\citet{aggarwal2017detecting} propose a decision-making model which considers both textual similarity and contextual similarity between bug reports.
By including contextual similarity in addition to textual similarity, they find that model performance improved significantly over using textual similarity alone.
They use cosine similarity to measure the similarity between features.
They construct a feature table with these similarity measures alongside other extracted features and train a decision tree classifier to predict whether or not a pair of bugs are duplicates.

\subsection{Detecting duplicate questions on Stack Overflow}

We also consider the task of detecting duplicate questions on Stack Overflow in our investigations, providing a more diverse overview of the task of duplicate detection in software engineering.
Like Bugzilla, Stack Overflow data is publicly available, making for a popular data set for natural language processing.
In a recent study, \citet{wang2020duplicate} investigated duplicate detection in Stack Overflow using deep learning.
Among the deep learning approaches that they considered are Convolutional Neural Networks (CNNs), Recurrent Neural Networks (RNNs) and Long Short-Term Memory (LSTM).
They consider a \textit{decision-making approach} as described in the Bugzilla task: that is, they consider pairs of questions and assign a label to each pair.
For generating sentence embeddings, they use Word2Vec-based average-pooling  \citep{mikolov2013word2vec}.
On Stack Overflow, metadata about questions is available, including the language that they involve.
Accordingly, \citet{wang2020duplicate} considers six different languages for the duplicate detection task.
Across all six languages, they find that Word2Vec with LSTM is the best-performing model.

\subsection{Detecting conflicting/duplicate requirements}

\citet{guo2021automatically} develop Finer Semantic Analysis-based Requirements Conflict detector (FSARC) for detecting conflicts in SRS documents. Requirements are parsed manually into an eight-tuple, forming a semantic meta-model.
The specific semantic elements are formulated based on heuristics, and mapping a requirement to the eight-tuple can be an ambiguous task~\citep{guo2021automatically}. 
Despite this limitation, their results show that the algorithm can achieve strong performance for conflict detection.
This approach is particularly limited in a few-shot setting as it requires manually mapping each requirement, which is often too time-consuming for few-shot learning~\citep{alex2021raft}.
\citet{malik2023transfer} propose the use of transformer models for conflict detection.
Their research focuses on a data-rich environment i.e., an environment in which a substantial amount of labeled data is available.
We extend their approach to a few-shot setting through the use of fine-tuned models, and also consider few-shot learning specific approaches such as PET and SetFit.

\subsection{Bug report entailment}

The process of taking in bug reports, labeling them, and assigning them to developers based on severity, ease of fixing, and other features---known as bug triaging---is well-studied in literature.
Several recent studies have found that accounting for the dependency between bugs can offer a substantial performance improvement to the triaging process~\citep{almhana2021considering, jahanshahi2022sdabt, jahanshahi2021dabt}.
Specifically, these studies found that by including bug dependencies when assigning reports to developers, the average amount of time it takes to solve a bug is decreased.
However, after an extensive literature review, we did not find any models which attempted to train a classifier to detect bug dependencies, neither in a few-shot nor a data rich environment.
Given the new-found importance of bug dependencies in the triaging process, we identify the process of training classifiers for dependency detection as a rich environment for new research.

\subsection{Few-shot learning methods}
Pattern-exploiting training (PET) is an important few-shot learning approach introduced by~\citet{schick2021exploiting}.
This approach to few-shot learning relies on access to a large amount of unlabeled data in addition to a small number of labeled examples and works by adding context to a problem through the use of a \textit{prompt}.
Specifically, \citet{schick2021exploiting} assume that a major hindrance to effective few-shot learning is the fact that it is often difficult to learn the task being solved when access to training data is limited.

As PLMs are trained using an MLM objective, they are already trained to handle the formulations generated by the PET algorithm.
As the pattern itself is able to provide additional information beyond the training data alone, it stands to reason that the combined knowledge of the PLM and the additional context added by the pattern can yield better results.
However, a single pattern and verbalizer does not always result in improved performance~\citep{schick2021exploiting, schick2021size}.
This is a result of having no access to a validation set to identify strong-performing patterns and verbalizers~\citep{schick2021exploiting}.
To remedy this, \citet{schick2021exploiting} propose training an ensemble of models on a set of pattern-verbalizer pairs.
Specifically, they create three pattern-verbalizer pairs, and for each pair they train three models using distinct random seeds for a total of nine models.

While training $N$ models on a handful of training instances can be relatively inexpensive for small $N$ (when compared to training a sequence classifier on a full-sized training dataset), both the storage space and the inference time increase $N$-fold, which is undesirable.
Accordingly, \citet{schick2021exploiting} derive a solution from \citet{hinton2015distilling}'s knowledge distillation approach which leverages access to an abundance of unlabeled data.
Specifically, a set of models $\setModel$ is generated by training each model on a unique pattern-verbalizer pair and random seed where the MLM loss function is given by the cross-entropy loss between $\Pr_{(\petPattern, \petVerbalizer)}(\instanceLabel\mid\petEx)$ and the one-hot encoded true label of the training instance, summed over the entire training set~\citep{schick2021exploiting}.
Then, the ensemble is used to generate soft-labels for the examples in the set of unlabeled data.
Specifically, for each unlabeled example $\petEx$, scores are computed via
\begin{equation}
    \petScore(\instanceLabel \mid \petEx)=
\dfrac{
    \sum_{\instanceModel\in\setModel} \petWeight(\instanceModel) \petScore(\instanceLabel\|\petPattern_\instanceModel(\petEx), \instanceModel)
}
{
    \sum_{\instanceModel\in\setModel} \petWeight(\instanceModel)
}
\end{equation}
where
$\petScore(\instanceLabel\|\petPattern_\instanceModel(\petEx), \instanceModel)$ gives the score assigned by the model $\instanceModel$ to the input $\petPattern_\instanceModel(\petEx)$, $\petPattern_\instanceModel$ is the pattern used for training the model $\instanceModel$,
and $\petWeight(\instanceModel)$ is the weight associated with the model $\instanceModel$.
The weight $\petWeight(\instanceModel)$ is equal to the accuracy of the model on the labeled training data \textit{before} training.
This weighting approach favours patterns that perform better even without training~\citep{schick2021exploiting}.
These scores are then converted to a probability distribution via softmax:
\begin{equation}
    \label{eq:pet-probability}
    \Pr_{(\petPattern, \petVerbalizer)}(\instanceLabel\mid\petEx)=\dfrac{\exp(\petScore(\instanceLabel \mid \petEx))}{\sum_{\instanceLabel'\in\setLabel} \exp(\petScore(\instanceLabel' \mid \petEx))}
\end{equation}
Finally, a sequence classifier is trained on the newly generated softly-labeled data and the labeled training data.

SetFit~\citep{tunstall2022efficient} is a recently introduced model which was shown to outperform PET on the RAFT leaderboard, a leaderboard designed to evaluate the performance of few-shot learning models~\citep{alex2021raft}.
SetFit relies on sentence transformers to train models. We briefly describe sentence transformers here.
As BERT and its variants can handle an input of two sentences, the procedure for finding the most similar pair of sentences in a set is straight-forward, but costly.
Specifically, this procedure requires comparing all sentence pairs.
For $n$ sentences, the number of comparisons to be made is $\mathcal{O}(n^2)$.
Accordingly, \citet{reimers2019sentence} propose Sentence-BERT (SBERT), a modification to BERT and BERT-like models which allows for the generation of strong vectors representing input sequences.
Similarity measures such as cosine similarity and clustering can then be used to find similar sentences.
These approaches allow SBERT to offer substantial performance improvements for sentence similarity tasks or sentence representation tasks.
SBERT derives a fixed-sized embedding by adding a mean pooling operation to the output of the BERT-like model~\citep{reimers2019sentence}.
While several loss can be used for training an SBERT model, for this research, only cosine similarity loss is used.
Specifically, given two sentences, the cosine similarity is computed and mean squared-error loss is employed with this similarity measure.

SetFit~\citep{tunstall2022efficient} uses contrastive learning to substantially increase the number of samples in a few-shot setting.
Specifically, given training data consisting of pairs of text and their label (i.e., $\setTrainData=\{(\petEx_i, \instanceLabel_i)\}$), we can create a binary classification dataset of triplets by taking pairs of examples from the dataset and assigning them the positive label ``1'' if they share the same class and assigning them the negative label ``0'' if they have different classes.
For each class label, $\setFitNumIterations$ positive and $\setFitNumIterations$ negative triplets are generated, and the union of all such triplets are used as the training data for contrastive learning~\citep{tunstall2022efficient}.

After the SBERT model is trained on the contrastive dataset, it is used to encode the original training data, thus generating a training set consisting of pairs of word embeddings and their class labels.
\citet{tunstall2022efficient} use a logistic regression model as the classification head for this training data.
When data is extremely scarce, the training procedure is extremely fast compared to the alternative few-shot learning methods.
Additionally, inference time is much quicker for SetFit models.
Finally, unlike for PET \citep{schick2021exploiting}, SetFit does not require manual input nor unlabeled data to train a model: given a training dataset $\setTrainData$, a model can immediately be trained without any task-specific programming.

\section{Methodology}\label{sec:methodology}
There are numerous studies into various text classification problems in the software engineering domain.
However, the research into few-shot learning applications to handle these tasks is limited.
In this section, we explore the applications of few-shot learning over a handful of software engineering tasks.
Specifically, we consider requirement conflict detection, bug duplicate detection, Stack Overflow duplicate question detection and bug entailment/dependency detection.

In the requirement conflict detection task, the objective is to identify conflicts between requirements in a software requirement specification (SRS) document.
This is accomplished by generating all possible pairings of requirements and assigning a label to each pairing.
The allowed labels are \textit{Neutral}, indicating that the pairing contains unrelated requirements and requires no manual intervention;
\textit{Duplicate}, indicating that the pairing contains two (almost) identical requirements that may be simultaneously satisfied;
and \textit{Conflict}, indicating that the pairing contains two related requirements which cannot be simultaneously satisfied.
A pairing labeled \textit{Conflict} must be addressed manually by revising one or both of the requirements in the pairing.
While \textit{Duplicate} requirements do not require manual intervention, they are of interest because if one of the requirements in the pairing is to change, they could become conflicting.
In addition, having \textit{Duplicate} requirements might lead to needlessly repeating some tasks.
Accordingly, it is important to consider the pairings labeled \textit{Duplicate} when adjusting requirements.

Bug duplicate detection is an important task in the bug-triaging process for two main reasons.
Firstly, in a triaging process where duplicates are not considered, each bug report in a pair of duplicates may be passed to two separate developers, resulting in wasted labour and potentially conflicts in the code created by the bug fixes introduced by each developer.
Secondly, a duplicate bug report may contain additional information, i.e., the pairing together can provide a more comprehensive overview of the problem.
Considering both reports together is therefore desirable.
In the bug duplicate detection task, we consider a collection of bug reports and generate all possible pairings of reports.
For each pairing we assign the label \textit{Duplicate}, indicating that the two reports pertain to the same issue, or \textit{Neutral}, indicating that the two reports are unrelated.

In recent literature, bug dependencies have become a popular topic for research in bug triaging~\citep{jahanshahi2022sdabt, jahanshahi2021dabt, almhana2021considering}.
Given two bug reports, \textA{} and \textB{}, \textA{} is said to depend on \textB{} if \textA{} cannot be resolved without resolving \textB{}.
Determining dependencies can ensure that \textB{} is prioritized so that \textA{} can be resolved, but it also offers an additional benefit.
It is logical to assign bugs that depend on one another to the same developer, as dependent bugs will often involve the same files~\citep{almhana2021considering}.
Incorporating these dependencies into the triaging process can therefore reduce the workload on the development team.
In the bug entailment task, we consider pairings $(\textA{}, \textB{})$ of bug reports.
We assign a pairing the label \textit{Entailment} if \textA{} depends on \textB{}, and the label \textit{Not Entailment} if \textA{} does not depend on \textB{}.

\subsection{Datasets}

To investigate the viability of few-shot learning methods in software engineering, we consider four tasks:
\begin{enumerate}
    \item detecting conflicts in software requirement specification (SRS) documents,
    \item detecting duplicate bug reports,
    \item detecting duplicate Stack Overflow questions, and
    \item detecting bug dependencies
\end{enumerate}
Note that unlike the generic text classification over individual sentences, each of these tasks involves classifying sentence pairs, $(\textA, \textB)$.
Table~\ref{tab:software-engineering-data-dist} provides summary statistics with respect to word counts for each dataset.
The summary statistics in Table~\ref{tab:software-engineering-data-dist} are given for the concatenation of $\textA$ and $\textB$, as the sentences are ultimately concatenated for each of the few-shot learning approaches employed.
As the number of tokens generated by a sentence depends on the tokenizer---and therefore, the underlying model checkpoint---we consider word counts to give a rough overview of the datasets as a whole.
This information is helpful when choosing the maximum sequence length used by the transformer models.

\begin{table}[htb]
\centering
\caption{Word counts for software engineering pair tasks}
\label{tab:software-engineering-data-dist}
\begin{tabular}{lcccc}
\toprule
Dataset &  Min. &       Mean &  Median &  Max. \\
\midrule
SRS Conflict  &   12 &  42~$\pm$~17 &      42 &   72 \\
Bugzilla Duplicate   &    6 &  34~$\pm$~17 &      34 &   70 \\
Stack Overflow Duplicate      &    6 &  28~$\pm$~13 &      28 &   51 \\
Bugzilla Entailment &    6 &  34~$\pm$~17 &      34 &   69 \\
\bottomrule
\end{tabular}
\end{table}

We consider training sets of size 25, 50, 100, 200, and 400 for all problems, and we use test sets with 2,000 examples and unlabeled datasets for use with PET of size 5,000.
For each task we repeat our experiments on three randomly drawn datasets and report average results.

\subsubsection{SRS Conflict Detection}

For the SRS conflict detection task, we make use of a proprietary dataset provided by IBM consisting of pairs of requirement specifications.
To each pair, we assign exactly one of three labels:
\textit{Conflict}, indicating that the two requirements cannot be simultaneously satisfied;
\textit{Duplicate}, indicating that while both requirements can be simultaneously satisfied, they are related in some way and if one is changed they may become conflicting;
and \textit{Neutral}, indicating that the requirement pairings are neither in conflict nor are they duplicates.
Table~\ref{tab:conflict-detection-examples} provides some examples from the conflict detection dataset to illustrate each of these labels.
From this brief overview of the dataset, we note that for the instance labeled \textit{Duplicate}, the words ``UAV'' and ``Hummingbird'' are treated synonymously.
The language model may have a difficult time noting this distinction with limited access to task-specific training data.

\begin{table}[htb]
\setlength{\tabcolsep}{15pt}
\renewcommand{\arraystretch}{2}
\centering
\caption{SRS conflict detection examples.}
\label{tab:conflict-detection-examples}
\resizebox{0.95\textwidth}{!}{
\begin{tabular}{p{0.4\linewidth}p{0.4\linewidth}l}
\toprule
Specification 1 & Specification 2 & Label \\
\midrule
The UAV shall instantaneously transmit information to the Pilot regarding mission-impacting failures. & The Hummingbird shall send the Pilot real-time information about malfunctions that impact the mission. & Duplicate \\
The UAV shall only accept commands from an authenticated Pilot. & The UAV shall accept commands from any Pilot controller. & Conflict \\
The UAV flight range shall be at least 20 miles from origin. & The UAV shall be able to transmit video feed to the Pilot and up to 4 separate UAV Viewer devices at once. & Neutral \\
\bottomrule
\end{tabular}
}
\end{table}

\subsubsection{Bugzilla Duplicate Detection}\label{sec:bugzilla-duplicate-task}
The first duplicate detection task that we consider is the Bugzilla duplicate detection task, which uses bugs scraped from Mozilla's Bugzilla\footnote{\url{https://bugzilla.mozilla.org}} using the REST API.\footnote{\url{https://bugzilla.mozilla.org/rest/bug}}
In this task, each sentence pair can be assigned one of two labels: \textit{Duplicate}, indicating that the two bug reports are duplicates and \textit{Neutral}, indicating that the bug reports are not duplicates.
Table~\ref{tab:bugzilla-fields} lists the queried fields that we used to construct our Bugzilla-based datasets.
For the duplicate detection task, we queried for bugs with a resolution of ``DUPLICATE'' to find duplicate bug reports.
For such bug reports, we queried for the ids listed in the \texttt{dupe\_of} field to get the duplicate bug reports.
We then joined these results to get duplicate pairs.
To create sentence pair data with the \textit{Neutral} label, we consider open bug reports.
As bug resolutions are handled by bug triagers, we collect bugs created no later than December 31st, 2021 to ensure that adequate time was afforded for bugs to be resolved as duplicates.
We collect the bugs in reverse-order from this date, ensuring that our data contains the most recent bug reports from this cutoff, and collect bugs from no earlier than January 1st, 2019.
When constructing our datasets, we ensure that for all sentence pairs $(\textA, \textB)$ in the training dataset, neither $\textA$ nor $\textB$ appears in any sentence pair in the test dataset.
\begin{table}[htb]
 \centering
 \caption{Bugzilla queried fields.}
 \label{tab:bugzilla-fields}
 \resizebox{0.72\textwidth}{!}{
 \begin{tabular}{ll}
 \toprule
 Field & Description \\
 \midrule
    \texttt{id}   & The primary key identifying this bug report. \\
    \texttt{summary}   & The title of the bug report which briefly summarizes it. \\
    \texttt{description}   & The detailed description of the bug report \\
    \texttt{creation\_time}   & The time at which the bug report was created. \\
    \texttt{resolution}   & The status of the bug. If blank, the bug is unresolved. \\
    \texttt{dupe\_of}   & A list of bug ids of which this bug is a duplicate. \\
    \texttt{depends\_on}   & A list of bug ids for other bug reports that this bug depends on.\\
\bottomrule
 \end{tabular}
 }
\end{table}

Table~\ref{tab:bugzilla-duplicate-examples} provides a few examples of from this data set.
\begin{table}[htb]
 \centering
 \caption{Bugzilla Duplicate Examples}
 \label{tab:bugzilla-duplicate-examples}
 \resizebox{0.8\textwidth}{!}{
 \begin{tabular}{p{.4\textwidth}p{.4\textwidth}l}
 \toprule
    Bug 1 & Bug 2 & Label \\
\midrule
Update preference gets changed to automatically install updates&Settings for “Firefox Update” get lost&Duplicate\\
Title tooltips flash off immediately&Title-text tooltip for an image appears and then instantly disappears again&Duplicate\\
allocation size overflow when overriding regexp exec to be dumb&Search : Few results are shown with Bing search engine depending on the region&Neutral\\
Count becomes zero after refreshing the page&Async/await all the things inside PersonalityProviderWorkerClass & Neutral\\
\bottomrule
 \end{tabular}
 }
\end{table}

 \subsubsection{Bugzilla Dependency Detection}

For the Bugzilla dependency detection task, also called the \textit{entailment} task, we consider a pair of bug reports $(\textA, \textB)$ and assign the label \textit{Entailment} if $\textA$ depends on $\textB$ and \textit{Not Entailment} if it does not.
 We follow a similar procedure to Section~\ref{sec:bugzilla-duplicate-task} when constructing datasets for this task.
 Specifically, we query for bugs over the same time period, we construct \textit{Neutral} pairs using the same strategy, and we ensure that no training data appears in our test dataset.
 For this task, we consider the \texttt{depends\_on} field from Table~\ref{tab:bugzilla-fields}, following the same procedure as was done for duplicate detection using the \texttt{dupe\_of} field.
 Table~\ref{tab:bugzilla-dependency-detection} provides some examples from the dependency detection dataset.
 
\begin{table}[htb]
 \centering
 \caption{Bug dependency data examples}
 \label{tab:bugzilla-dependency-detection}
 \resizebox{0.8\textwidth}{!}{
 \begin{tabular}{p{.4\textwidth}p{.4\textwidth}l}
 \toprule
    Bug 1 & Bug 2 & Label \\
\midrule
    {[meta]} Oblivious DoH support & Consider to add confirmation mechanism for ODoH & Entailment \\
    Let the conversations reorder and save the order & Convert conversation list to tree-listbox & Entailment  \\
    Remove boilerplate necessary to introduce system metric media queries. & Browser.getVersion to return 1.3 & Not Entailment \\ 
    Array.sort is not working properly with return 0 & Fix various compile warnings in NSS & Not Entailment\\
\bottomrule
 \end{tabular}
 }
\end{table}
 
\subsubsection{Stack Overflow Duplicate Detection}
Finally, for the Stack Overflow duplicate detection task, we query the Stack Exchange Data Explorer for questions from Stack Overflow.\footnote{https://data.stackexchange.com/stackoverflow/queries}
In this task, as for Bugzilla duplicate detection, we assign each sentence pair one of the labels \textit{Duplicate} and \textit{Neutral}.
As posts on Stack Overflow generally have a tag indicating the applicable language/concepts, we selected only questions that had the tag \textit{Python}.
As in practice cross-language questions would not be labeled as duplicates, selecting only Python questions would likely be done in a real application of this duplicate detection task.
Querying for the Stack Overflow questions involves joining a considerable number of tables.
Table~\ref{tab:stack-overflow-duplicate-examples} provides some example sentence pairs and their labels for this dataset.
As for the Bugzilla detection dataset, we consider open posts (i.e., \textit{Neutral} pairs) from no later than December 31st, 2021 to ensure an adequate amount of time for questions to be closed as duplicates.

\begin{table}[htb]
 \centering
 \caption{Stack Overflow Duplicate Examples}
 \label{tab:stack-overflow-duplicate-examples}
 \resizebox{0.8\textwidth}{!}{
 \begin{tabular}{p{.4\textwidth}p{.4\textwidth}l}
 \toprule
    Title 1 & Title 2 & Label \\
\midrule
    IMAP4: How to correctly decode UTF-8 encoded message body? & Python email quoted-printable encoding problem & Duplicate \\ 
Jinja2: how to evaluate Python variable in an if-statement? & Jinja2 template evaluate variable as attribute, & Duplicate \\ 
    python csv read column function not giving correct output & WSGIServer not working with https and python3 & Neutral \\
select and filtered files in directory with enumerating in loop & How to set any polygon the same total width & Neutral \\ 
\bottomrule
 \end{tabular}
 }
\end{table}

For each of the Bugzilla and Stack Overflow-based tasks, a preliminary analysis revealed that using the entire question body had a negative impact on experimental results.
As these question bodies are often substantially longer than the maximum sequence length that transformer models are trained on, they generally have to be truncated significantly.
For algorithms like PET that introduce a mask token and additional text, the question/bug bodies are truncated even further.
Accordingly, for the Bugzilla dependency detection, Bugzilla duplicate detection, and Stack Overflow duplicate detection tasks, we consider only the question titles/bug summaries.

\subsection{Adaptation of SetFit to software engineering tasks}

The contrastive-learning premise for SetFit naturally lends it to classifying individual sentences.
However, for our research into the applications of few-shot learning in software engineering, we require models to handle sentence pairs.
Accordingly, we extend SetFit to the sentence pair classification task by joining the sentence pair with a separator token, ``\segsep{}'', and applying segment embeddings to each sentence, borrowing the same mechanism which is used for sequence classifier sentence pair classification.

\subsection{Adaptation of PET to software engineering tasks}

For the software engineering tasks, which all involve sentence pairs, we denote the first sentence by \textA{} and the second by \textB{}.
For the MLM objective, a mask token must be inserted into the pattern.
We denote the mask token in patterns by \mask{} and the barrier between two segments of text in a pattern by ``\segsep{}''.

\paragraph{Bugzilla Bug Dependency Detection Task.} For the Bugzilla bug dependency/entailment task, we consider the three patterns used by \citet{schick2021exploiting} on their entailment task:
\begin{enumerate}
\item ``\textB{}'' ?\segsep{} \mask{} , ``\textA{}''
\item \textB{} ?\segsep{} \mask{} , \textA{}
\item ``\textB{}'' ?\segsep{} \mask{} . ``\textA{}''
\end{enumerate}
We use the same verbalizer for each pattern:
\begin{align*}
    \petVerbalizer(\text{Not Entailment})&=\text{No}\\
    \petVerbalizer(\text{Entailment})&=\text{Yes}
\end{align*}
Using common words in our verbalizer as suggested by \citet{schick2020petal}.

\paragraph{Stack Overflow Duplicate Detection Task.}
For the Stack Overflow duplicate detection task, we consider the following three patterns:
\begin{enumerate}
\item ``\textB{}''? \segsep{} \mask{}. ``\textA{}''.
\item Are ``\textA{}'' and ``\textB{}'' the same question? \mask{} .
\item Are ``\textA{}'' and ``\textB{}'' duplicates? \mask{} .
\end{enumerate}
We use the same verbalizer for each pattern:
\begin{align*}
    \petVerbalizer(\text{Neutral})&=\text{No}\\
    \petVerbalizer(\text{Duplicate})&=\text{Yes}
\end{align*}

\paragraph{Bugzilla Duplicate Detection Task.}
For the Bugzilla duplicate detection task, we consider the following three patterns:
\begin{enumerate}
\item ``\textB{}''? \segsep{} \mask{}. ``\textA{}''.
\item Are ``\textA{}'' and ``\textB{}'' the same problem? \mask{} .
\item Are ``\textA{}'' and ``\textB{}'' duplicates? \mask{} .
\end{enumerate}
We use the same verbalizer for each pattern:
\begin{align*}
    \petVerbalizer(\text{Neutral})&=\text{No}\\
    \petVerbalizer(\text{Duplicate})&=\text{Yes}
\end{align*}

\paragraph{Conflict Detection Task}
For the Conflict detection task, we use the following three patterns:
\begin{enumerate}
    \item ``\textA{}''? \segsep{} \mask{}, ``\textB{}''.
    \item Given ``\textA{}'', we can conclude that ``\textB{}'' is \mask{}.
    \item ``\textA{}'' means ``\textB{}''.\segsep{} \mask{}.
\end{enumerate}
We use a different verbalizer for each pattern.
For Pattern 1, we use
\begin{align*}
    \petVerbalizer(\text{Neutral})&=\text{Maybe}\\
    \petVerbalizer(\text{Duplicate})&=\text{Yes}\\
    \petVerbalizer(\text{Conflict})&=\text{No}
\end{align*}
For Pattern 2, we use
\begin{align*}
    \petVerbalizer(\text{Neutral})&=\text{neither}\\
    \petVerbalizer(\text{Duplicate})&=\text{true}\\
    \petVerbalizer(\text{Conflict})&=\text{false}
\end{align*}
and for Pattern 3, we use
\begin{align*}
    \petVerbalizer(\text{Neutral})&=\text{Neither}\\
    \petVerbalizer(\text{Duplicate})&=\text{True}\\
    \petVerbalizer(\text{Conflict})&=\text{False}
\end{align*}

\subsection{Experimental Setup}

As outlined by~\citet{perez2021true}, tuning hyperparameters on validation data can have a substantial performance impact on experimental results.
Accordingly, we rely heavily on the literature, particularly the works of~\citet{schick2021exploiting}, \citet{tam2021adapet}, \citet{tunstall2022efficient}, and \citet{zhang2021revisiting}, for choosing hyperparameters, with one exception.
In our experimentation, we found that training SetFit models with the same hyperparameters was not robust against a change in model checkpoint. 
Accordingly, we considered training each SetFit model for 1 and 3 epochs, and found that \bertLarge{} and \robertaLarge{} yielded stronger results when using a single epoch, but \debertaLarge{} required 3 training epochs.
Without this hyperparameter sweep, model performance was degraded substantially.

Aside from this exception, the selected hyperparameters do not vary for each of the algorithms.
For few-shot datasets, we use 1,000 training steps for PET and fine-tuning.
For all approaches, we employ a batch size of 16, and a learning rate of $10^{-5}$, and a maximum sequence length of 256.
For training the PET sequence classifier, we use a temperature of 2 for generating soft labels, and train the model for 5,000 steps on the softly labeled data.
For full-sized datasets we use a batch size of 16 and train all models for 5,000 training steps using a learning rate of $10^{-5}$.
We do not conduct hyperparameter tuning on the full-sized datasets.

We implement fine-tuning using the torch-based checkpoints from the \texttt{transformers} library.\footnote{\url{https://github.com/huggingface/transformers}}
We use the implementation of SetFit provided by \citet{tunstall2022efficient}.\footnote{\url{https://github.com/huggingface/setfit}}
For PET, we modify the implementation provided by \citet{schick2021exploiting}\footnote{\url{https://github.com/timoschick/pet}} to allow for distributed computing, and introduce our hand-crafted task-specific patterns and verbalizers.

We conduct our experiments using the resources provided by the Digital Research Alliance of Canada (DRAC).\footnote{\url{https://alliancecan.ca/}}
For each of the considered approaches---fine-tuning, PET, and SetFit---we use GPUs for improved training and inference speed.
For some approaches (particularly those using \debertaLarge{}), memory usage was high, requiring the usage of GPUs with 32GB of memory.
For approaches using \bertLarge{}, 16GB was often sufficient.
Accordingly, these experiments are \emph{not} run on identical hardware.
In many cases, we also use gradient accumulation over more than one step (while maintaining an effective batch size of 16 as stated above).

\section{Results}\label{sec:results}
This section details our results for each of the focused checkpoints on the four tasks.
We begin by comparing the performance of each model on full-sized training datasets, and then compare their performance in a few-shot setting.
Finally, we investigate their performance with respect to increasing training set size.

\subsection{Performance Analysis of Transformer Models}

Table~\ref{tab:eng-full-sized} provides the accuracy attained by each of the model checkpoints using vanilla fine tuning for 5,000 training steps on the full-sized datasets.
Across each task, we observe the \debertaLarge{} model performing the best, followed by \robertaLarge{}, followed by \bertLarge{}.
This is expected as the three models are a set of incremental improvements on \bertLarge{}, with the \debertaLarge{} model being the most recent iteration.
\begin{table}[htb]
\centering
\caption{Full-sized dataset results for each model.}
\label{tab:eng-full-sized}
\begin{tabular}{lcccc}
\toprule
&        Bugzilla &        Conflict &      Entailment &  Stack Overflow \\
\midrule
Num. Examples &            4,400 &         3,626 & 10,949 & 7,965  \\
\midrule
\bertLarge{}    &  $95.1 \pm 0.6$ &  $87.3 \pm 1.3$ &  $91.6 \pm 0.3$ &  $92.9 \pm 0.4$ \\
\robertaLarge{} &  $95.5 \pm 0.2$ &  $89.0 \pm 1.0$ &  $92.8 \pm 0.3$ &  $94.0 \pm 0.6$ \\
\debertaLarge{} &  $\mathbf{96.0 \pm 0.2}$ &  $\mathbf{90.1 \pm 1.1}$ &  $\mathbf{94.6 \pm 0.3}$ &  $\mathbf{96.0 \pm 0.2}$ \\
\bottomrule
\end{tabular}
\begin{tablenotes}
    \centering
    \footnotesize
    \item \textbf{Bold:}~Best score for a task.
\end{tablenotes}
\end{table}

Table~\ref{tab:eng-few-shot-50} provides summary results of few-shot learning using 50 labeled examples.
We observe that, unlike the full-sized results, \debertaLarge{} is no longer the de facto best model, as it only attains the highest accuracy for the SRS conflict detection dataset.
Despite performing the worst in the full-sized dataset results, \bertLarge{} attains the highest accuracy on two of the four datasets using vanilla fine-tuning, beating out the next best accuracy by 4.6\% and 2.9\% for the Bugzilla and Entailment datasets, respectively.
The results show that, for a fixed model checkpoint, PET tends to be the best performing approach, with the exception of \bertLarge{}, where fine-tuning tends to be the best performer.
When considering the best performing algorithm for each model (shown with an underline) we see some substantial performance gains: for example, choosing \debertaLarge{} PET over \bertLarge{} fine-tuning for the Bugzilla dupicate detection dataset results in a drop of 10\% in accuracy.
We also observe that SetFit performs its best for the \debertaLarge{} model, having considerably lower accuracy for the \bertLarge{} and \robertaLarge{} models when compared to the other methods using the same checkpoint.

\renewcommand{\arraystretch}{1.05}
\begin{table}[htb]
\centering
\caption{Few-shot results for 50 labeled examples.}
\label{tab:eng-few-shot-50}
\begin{tabular}{llcccc}
\toprule
                &  &      Bugzilla &      Conflict &    Entailment & Stack Overflow \\
\midrule
\bertLarge{} & Fine-tune &  \underline{$\mathbf{90.7\pm1.4}$} &  $75.2\pm1.1$ &  \underline{$\mathbf{84.6\pm2.8}$} &   \underline{$85.8\pm1.7$} \\
                & PET &  $78.9\pm4.0$ &  \underline{$76.7\pm1.8$} &  $76.6\pm3.4$ &   $77.4\pm4.8$ \\
                & SetFit &  $75.8\pm5.3$ &  $67.7\pm4.5$ &  $70.3\pm1.6$ &   $73.5\pm3.2$ \\
\midrule                
\robertaLarge{} & Fine-tune &  $80.7\pm7.6$ &  $77.8\pm2.5$ &  $74.7\pm4.7$ &   $75.0\pm5.7$ \\
                & PET &  \underline{$86.1\pm2.2$} &  \underline{$79.7\pm0.8$} &  \underline{$81.7\pm2.2$} &   \underline{$\mathbf{86.5\pm3.4}$} \\
                & SetFit &  $74.3\pm3.2$ &  $68.1\pm7.5$ &  $71.7\pm4.2$ &   $70.1\pm6.4$ \\
\midrule  
\debertaLarge{} & Fine-tune &  $79.4\pm4.1$ &  $74.8\pm4.1$ &  $74.9\pm4.2$ &   $80.0\pm3.0$ \\
                & PET &  \underline{$80.7\pm1.9$} &  \underline{$\mathbf{81.3\pm2.4}$} &  \underline{$80.1\pm1.7$} &   \underline{$84.1\pm8.3$} \\
                & SetFit &  $77.9\pm5.4$ &  $78.5\pm2.9$ &  $75.3\pm5.0$ &   $81.4\pm5.6$ \\
\bottomrule
\end{tabular}
\begin{tablenotes}
    \centering
    \footnotesize
    \item \textbf{Bold:}~Best score for a task.
    \item \underline{Underline:}~Best score for a fixed checkpoint on a task.
\end{tablenotes}
\end{table}

\subsection{Impact of Training Set Size}

In this subsection, we investigate the performance of each few-shot learning approach for the three considered model checkpoints with respect to increasing training set size.
We consider training sets of size 25, 50, 100, 200, and 400 and plot their accuracy, weighted average F1 score, and macro average F1 score.
We use a $\log_2$ scale for the $x$-axis in each of our plots for clarity, meaning that each point on the plot represents a doubling in the amount of training data.

\begin{figure}[!htb]
\centering
\subfloat[\bertLarge{} (Acc)\label{fig:bugzilla-bert-large-uncased-accuracy}]{\includegraphics[width=0.3\textwidth]{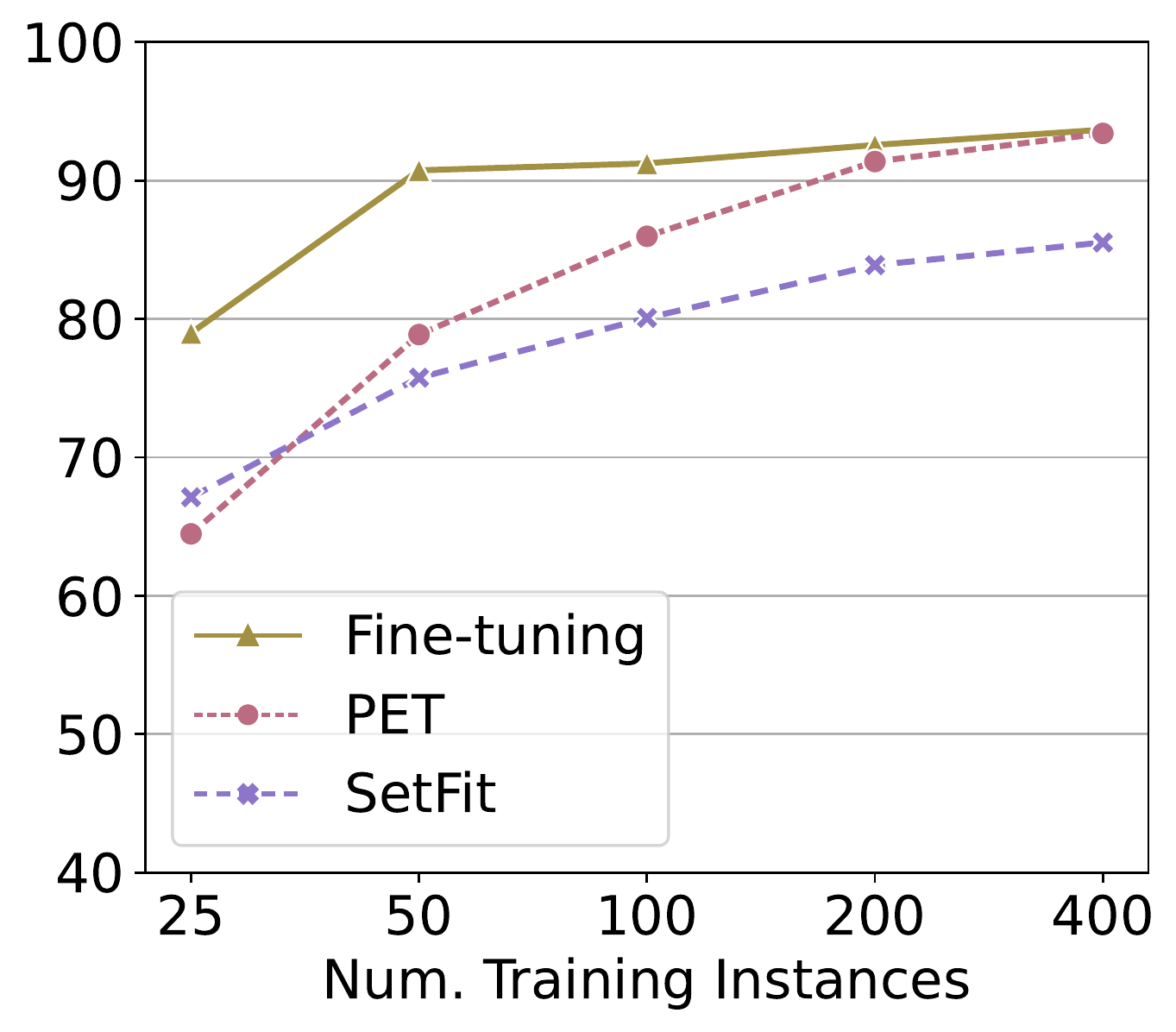}}
\subfloat[\robertaLarge{} (Acc)\label{fig:bugzilla-roberta-large-accuracy}]{\includegraphics[width=0.3\textwidth]{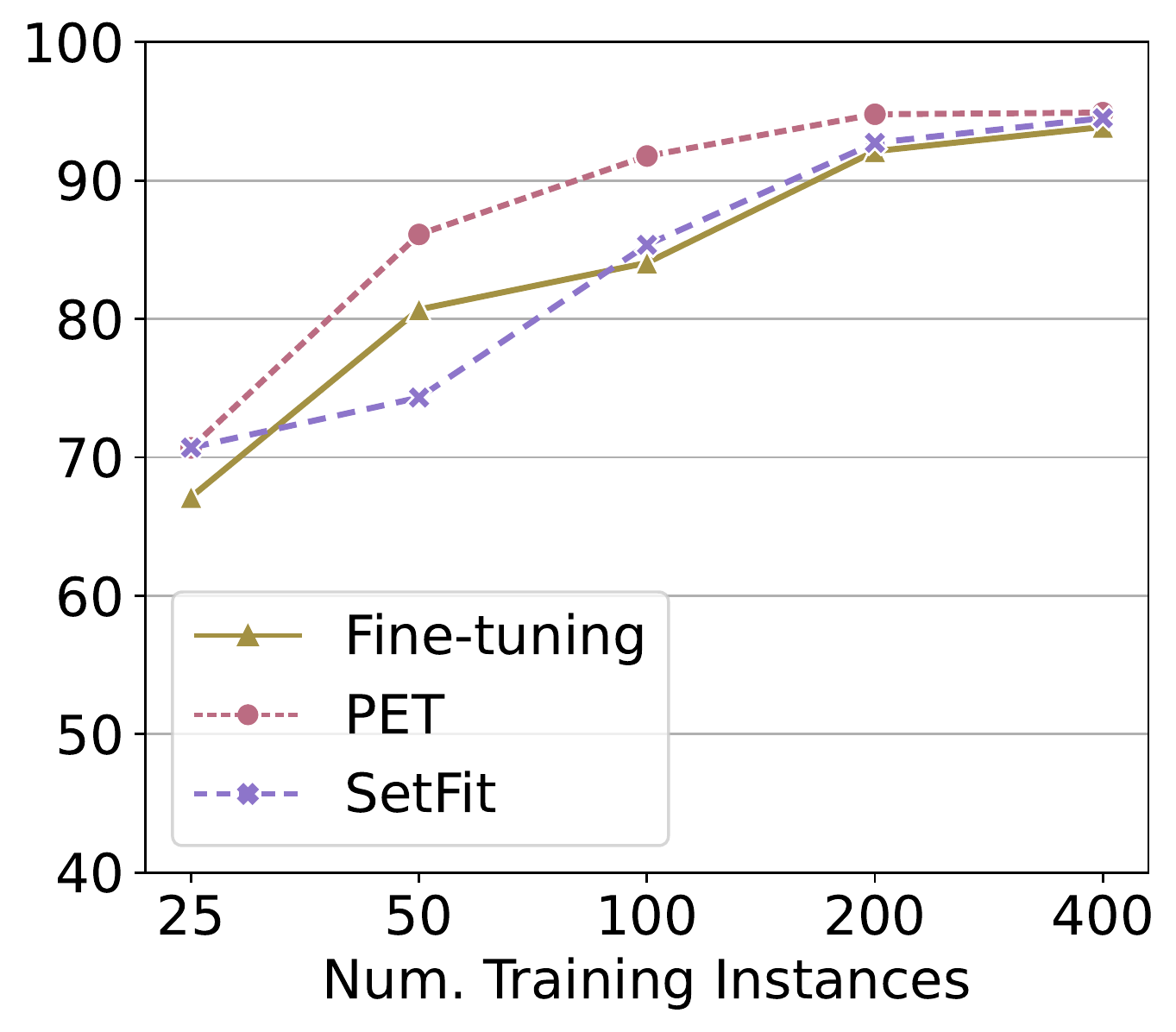}}
\subfloat[\debertaLarge{} (Acc)\label{fig:bugzilla-deberta-v3-large-accuracy}]{\includegraphics[width=0.3\textwidth]{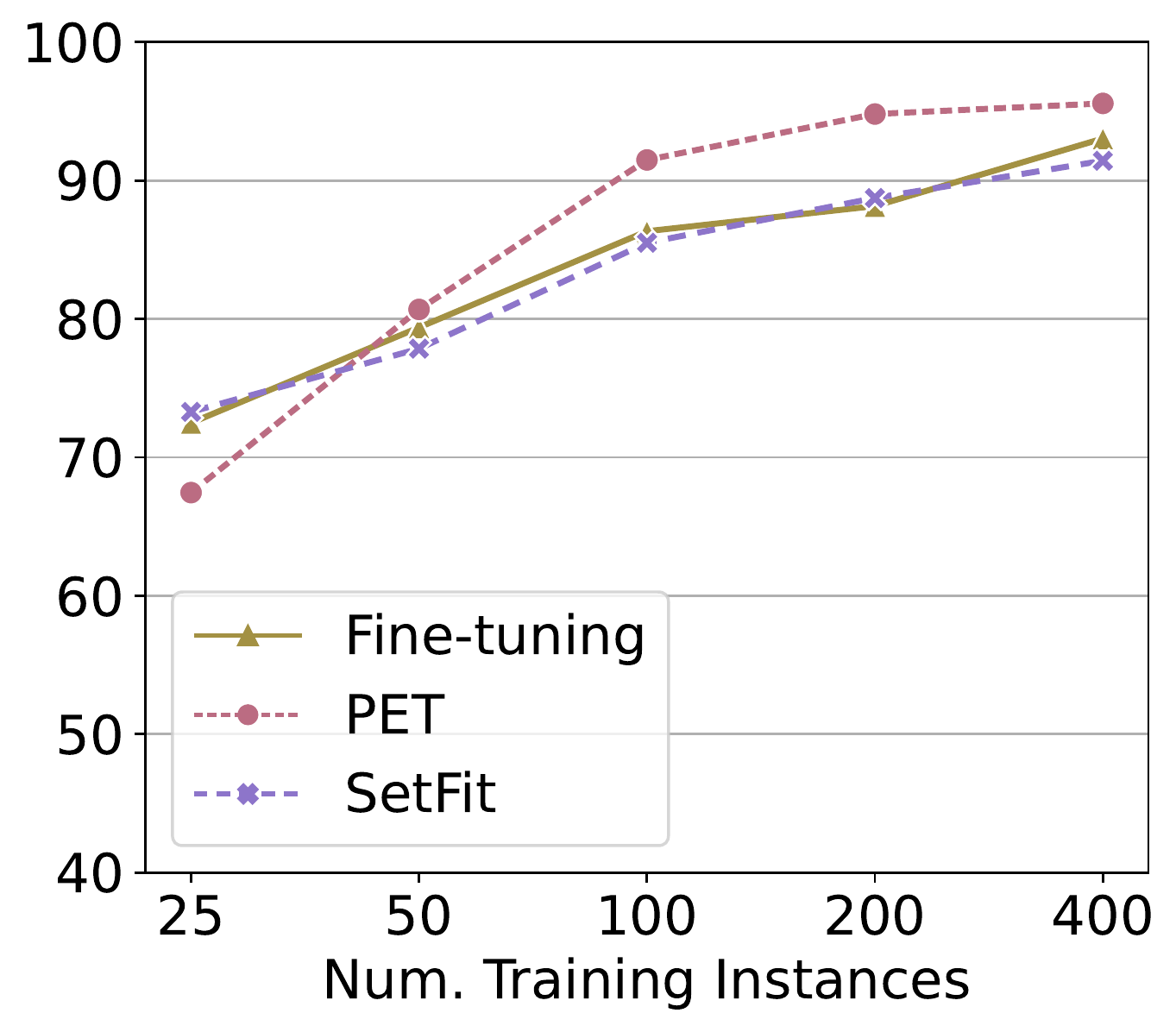}}
\\
\subfloat[\bertLarge{} (WF1)\label{fig:bugzilla-bert-large-uncased-weighted-avg-f1-score}]{\includegraphics[width=0.3\textwidth]{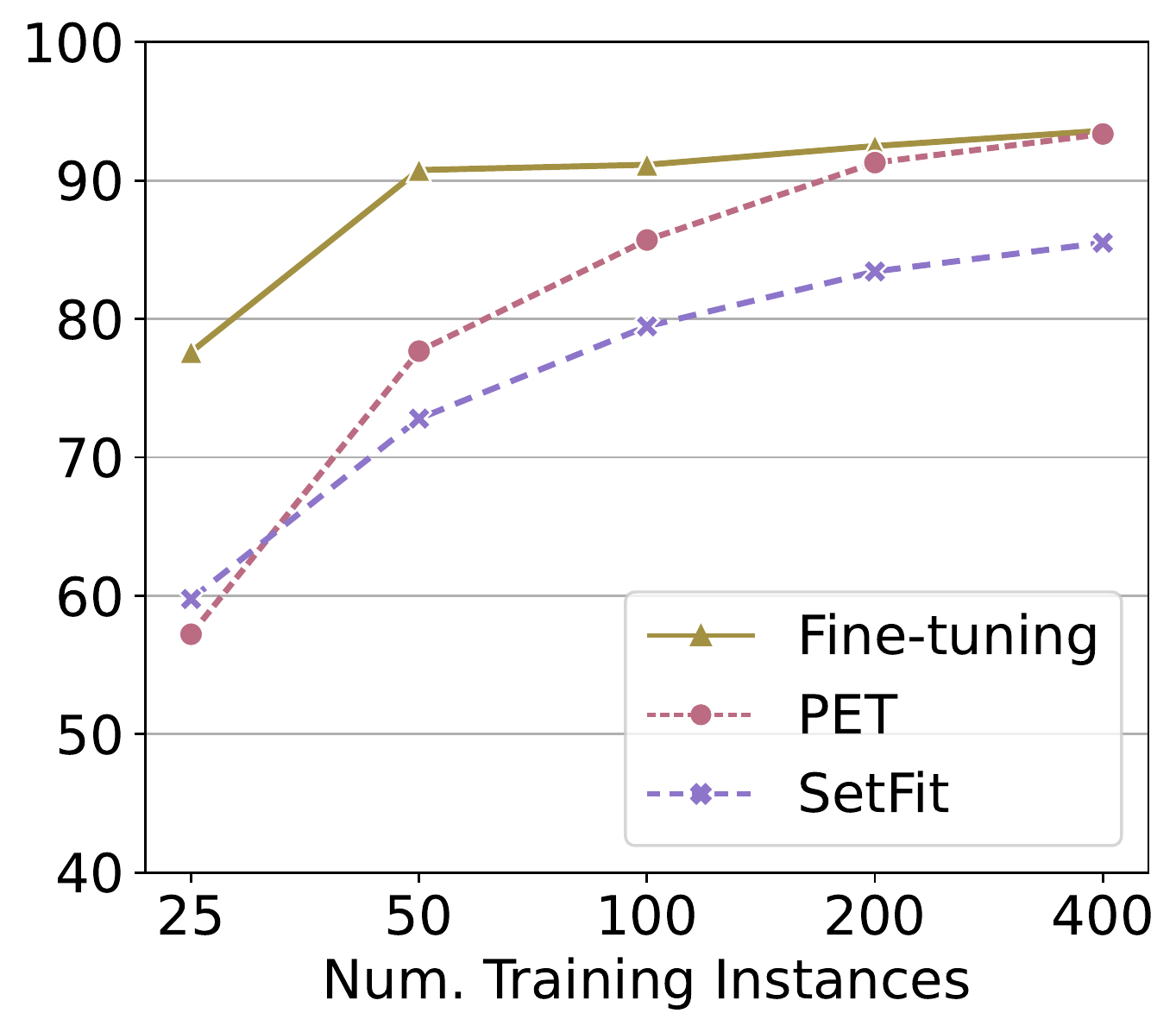}}
\subfloat[\robertaLarge{} (WF1)\label{fig:bugzilla-roberta-large-weighted-avg-f1-score}]{\includegraphics[width=0.3\textwidth]{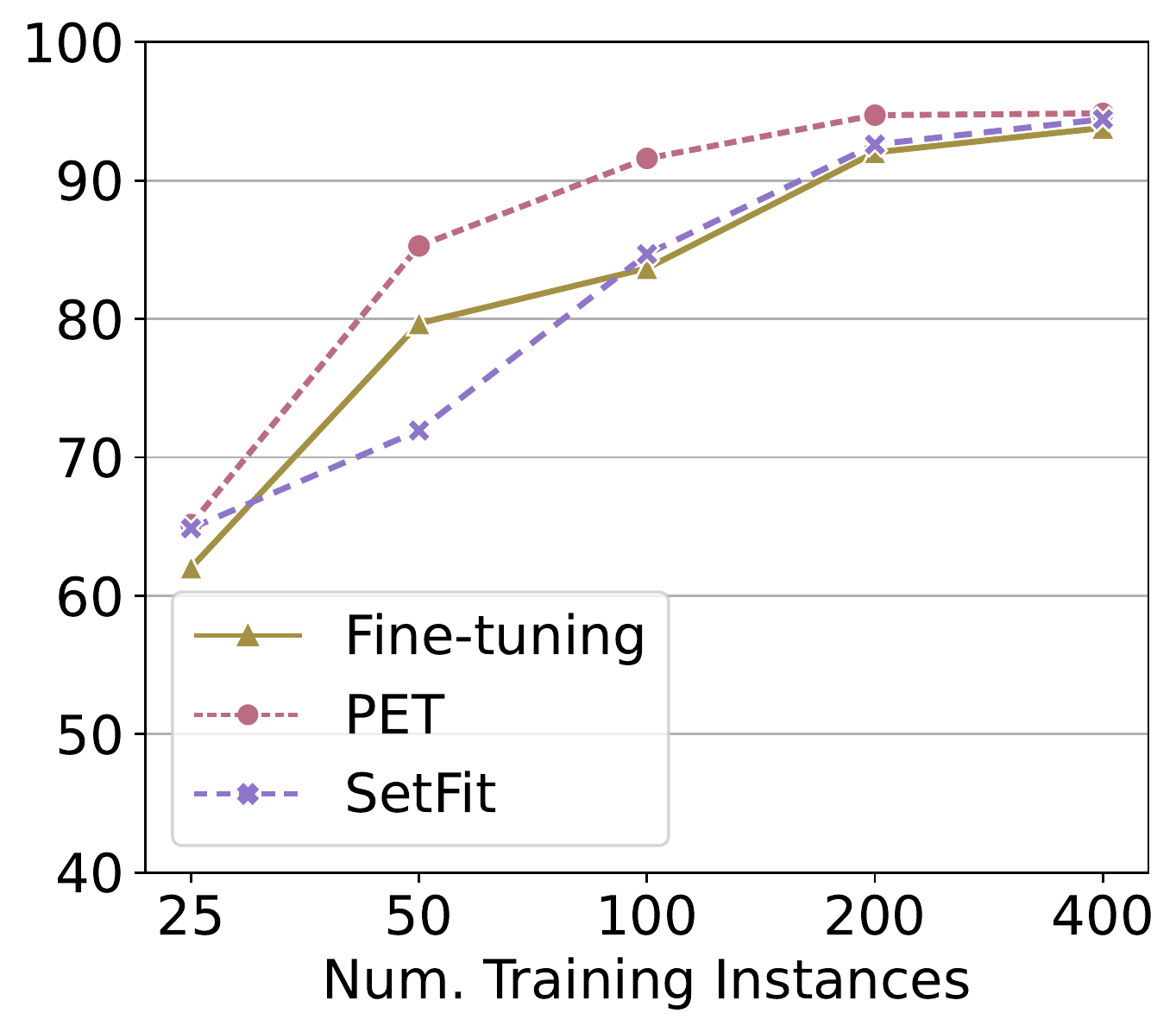}}
\subfloat[\debertaLarge{} (WF1)\label{fig:bugzilla-deberta-v3-large-weighted-avg-f1-score}]{\includegraphics[width=0.3\textwidth]{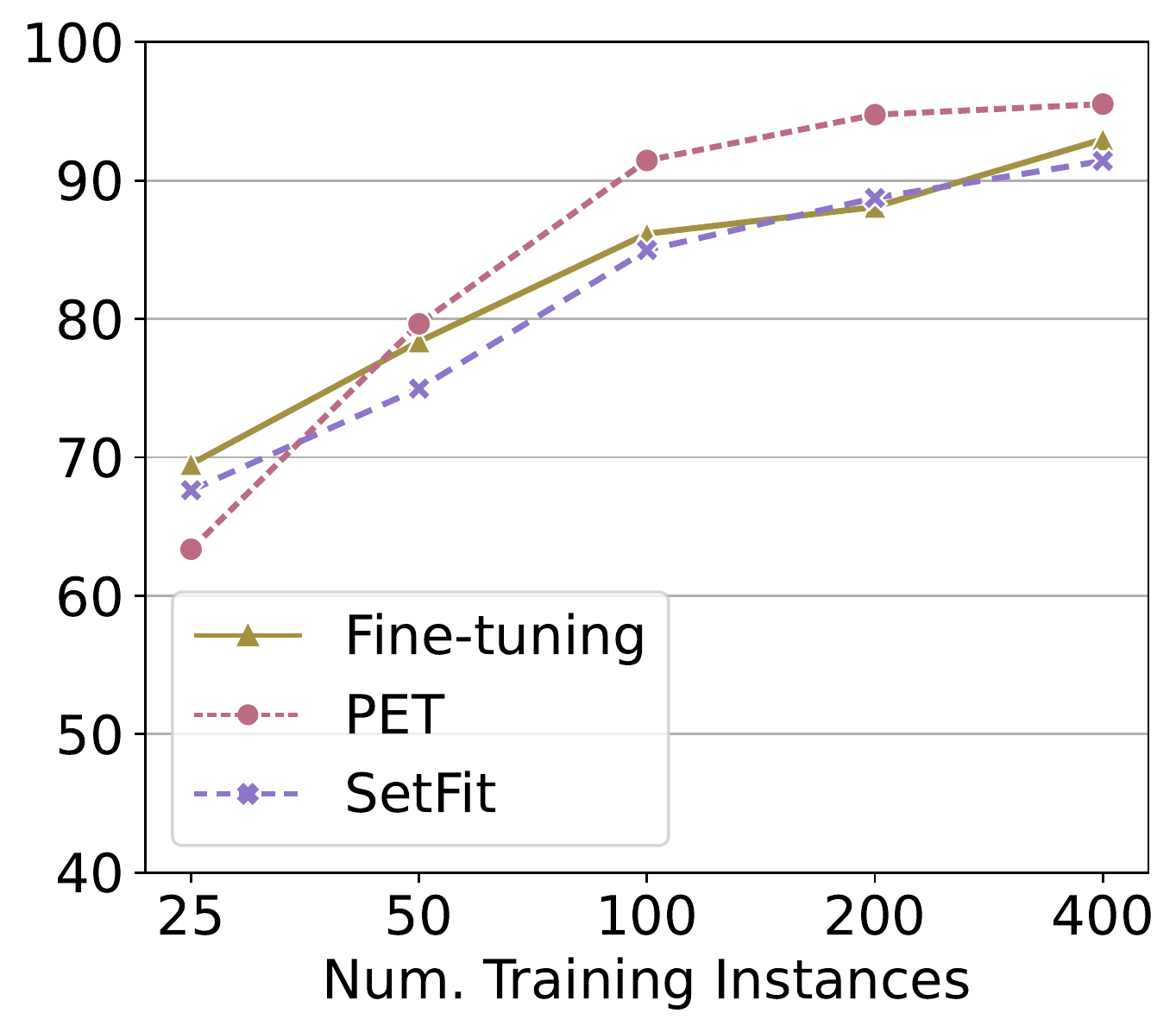}}
\\
\subfloat[\bertLarge{} (MF1)\label{fig:bugzilla-bert-large-uncased-macro-avg-f1-score}]{\includegraphics[width=0.3\textwidth]{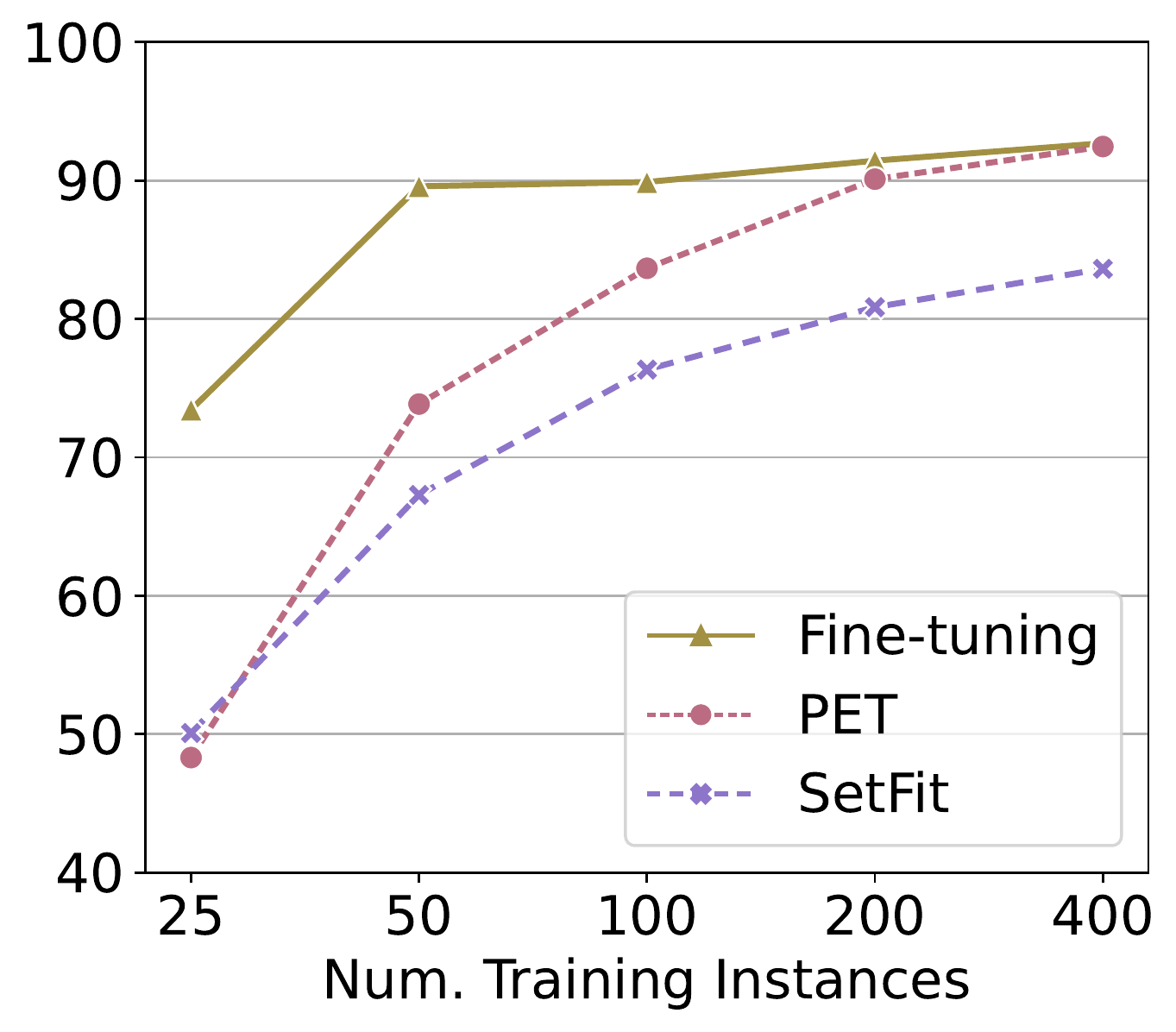}}
\subfloat[\robertaLarge{} (MF1)\label{fig:bugzilla-roberta-large-macro-avg-f1-score}]{\includegraphics[width=0.3\textwidth]{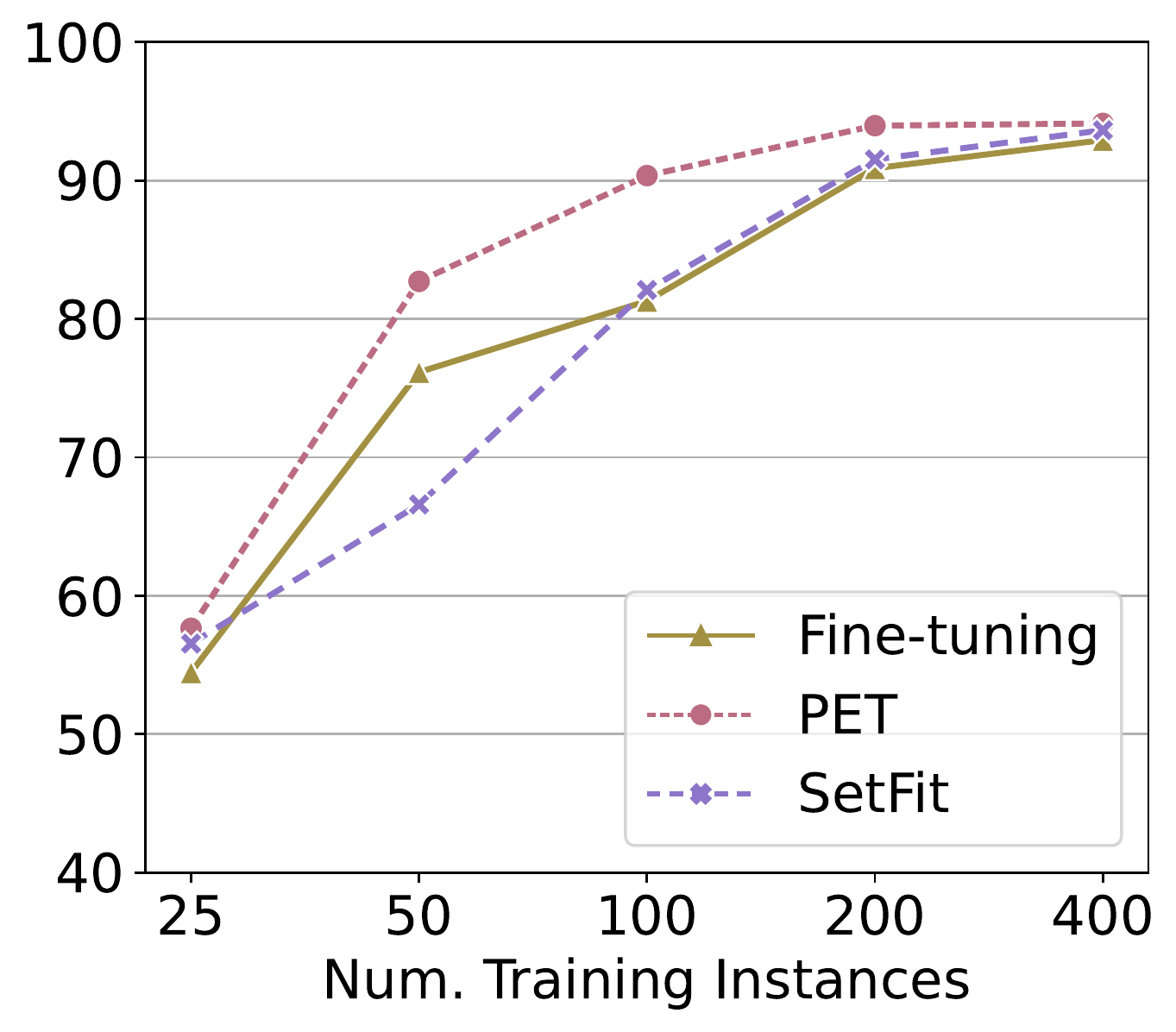}}
\subfloat[\debertaLarge{} (MF1)\label{fig:bugzilla-deberta-v3-large-macro-avg-f1-score}]{\includegraphics[width=0.3\textwidth]{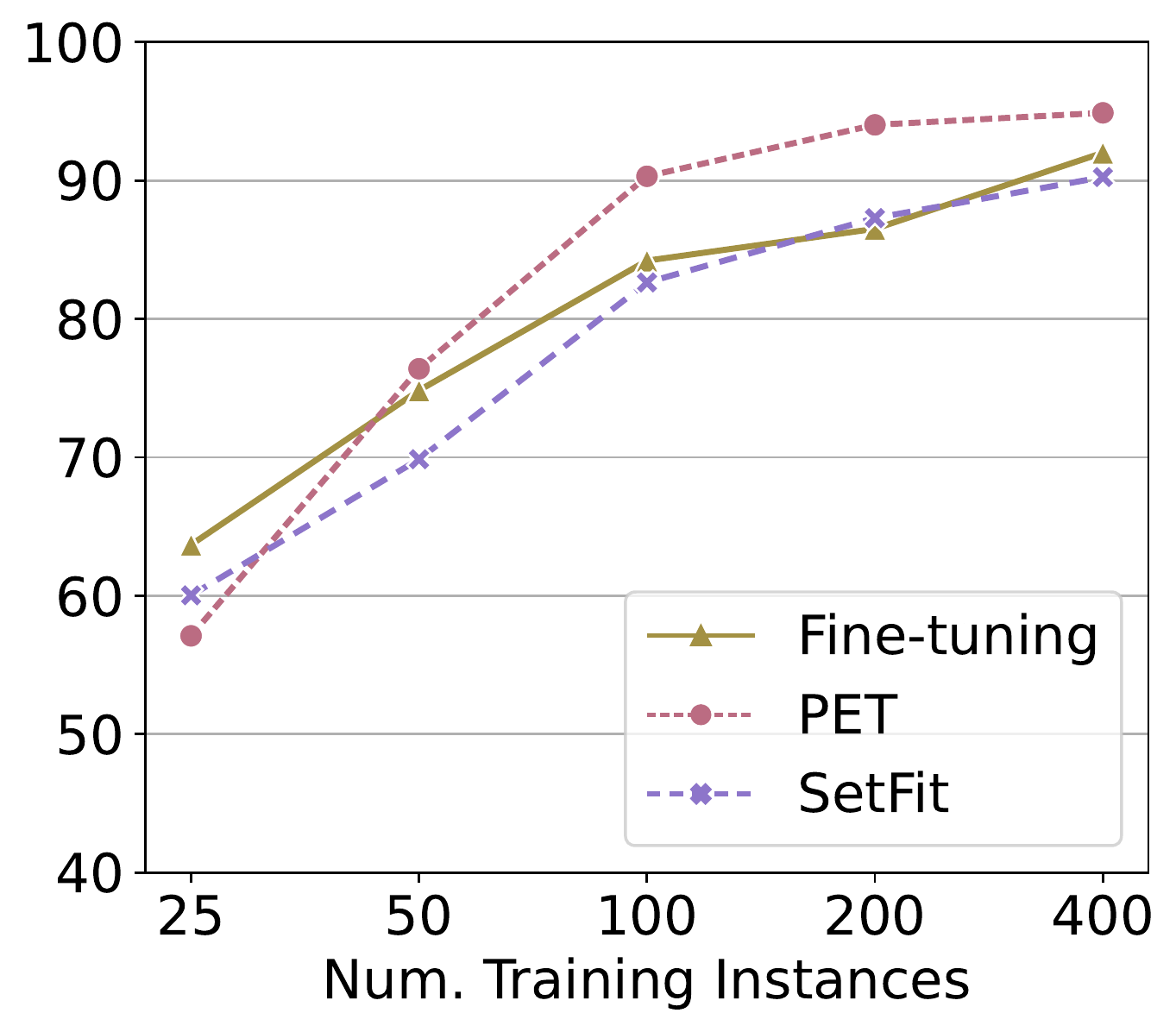}}
\\
\caption{Bugzilla duplicate detection results.}
\begin{tablenotes}
\centering
\footnotesize
\item Acc: Accuracy
\item WF1: Weighted Average F1-Score
\item MF1: Macro Average F1-Score
\end{tablenotes}
\label{fig:bugzilla-all-results}
\end{figure}

Figure~\ref{fig:bugzilla-all-results} shows the performance of each algorithm on the Bugzilla dataset.
The results for \bertLarge{} further illustrate the performance of fine-tuning outlined in Table~\ref{tab:eng-few-shot-50}.
It is not until 200 data instances that the performance of PET begins to match that of fine-tuning.
Notably, fine-tuning with \bertLarge{} nears its maximum performance for the considered datasets after just 50 training examples.
PET with 100 training instances using \robertaLarge{} or \debertaLarge{} is competitive with fine-tuning using \bertLarge{}.
We observe that fine-tuning has the largest performance drop when changing model checkpoints, particularly when data access is limited: moving from \bertLarge{} to \robertaLarge{}, the accuracy of fine-tuning decreases by more than 10\% when using 25 training instances.
In contrast, when switching to \robertaLarge{}, the accuracy of PET increases by about 5\%.
These observations and the rest of the results in Figure~\ref{fig:bugzilla-all-results} suggest that while vanilla fine-tuning with \bertLarge{} may be a good option for few-shot learning, this checkpoint is not a good base MLM model for PET for this task.
Finally, we note that for 25 training instances, SetFit tends to perform similarly to PET but for larger training set sizes, the improved training speed of SetFit is probably not worth the significant decrease in model accuracy.

\begin{figure}[!htb]
\centering
\subfloat[\bertLarge{} (Acc)\label{fig:entailment-bert-large-uncased-accuracy}]{\includegraphics[width=0.3\textwidth]{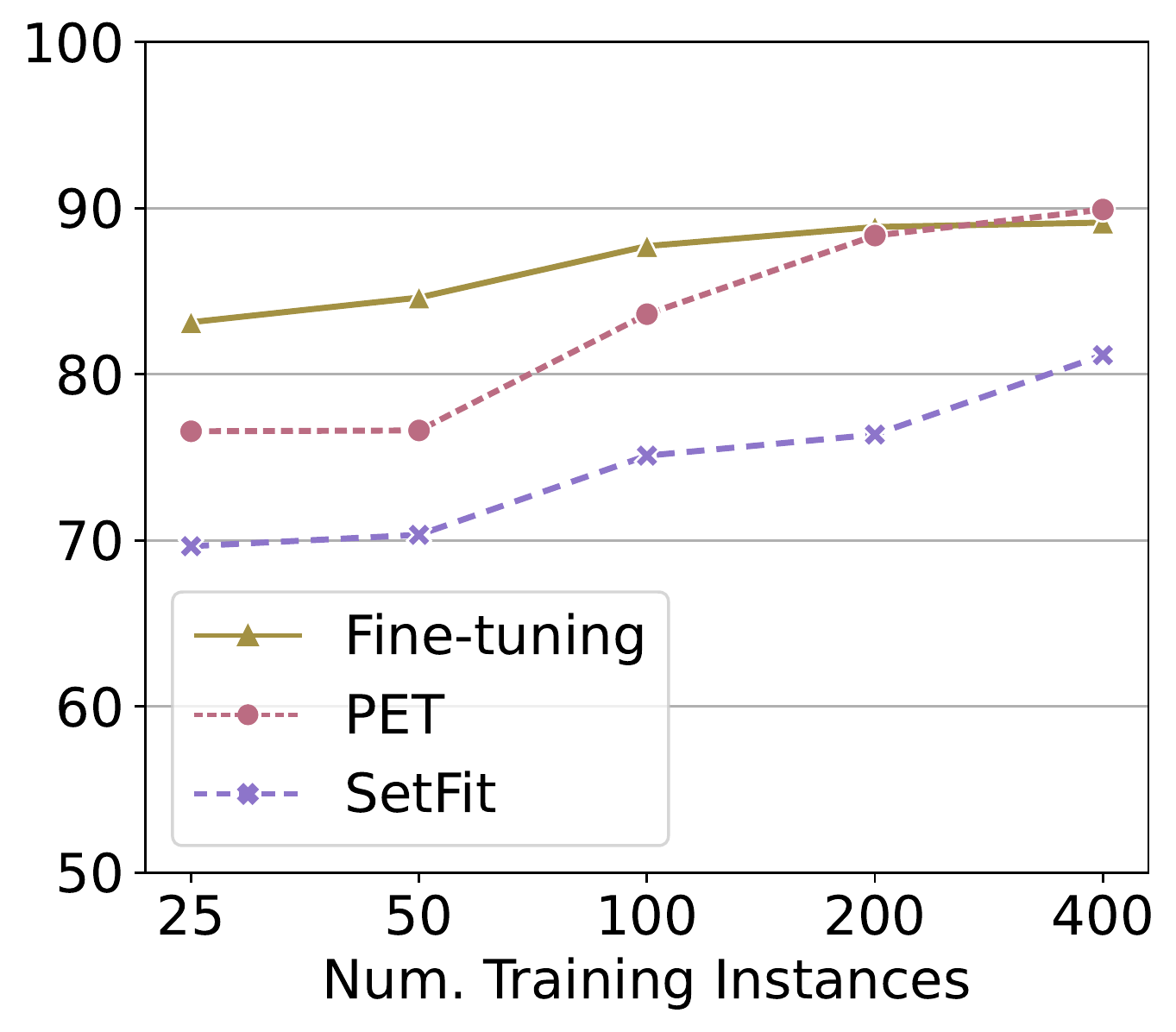}}
\subfloat[\robertaLarge{} (Acc)\label{fig:entailment-roberta-large-accuracy}]{\includegraphics[width=0.3\textwidth]{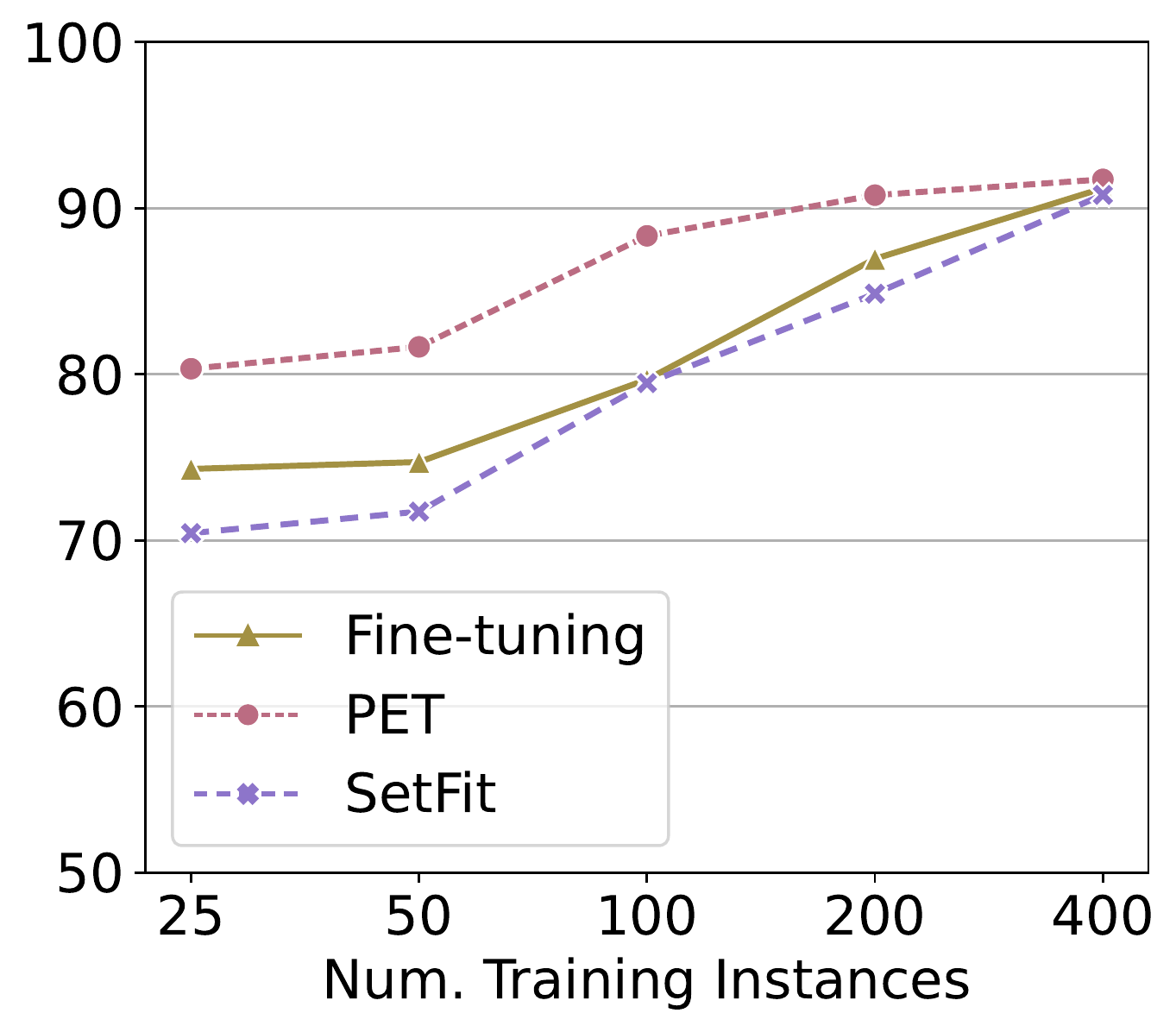}}
\subfloat[\debertaLarge{} (Acc)\label{fig:entailment-deberta-v3-large-accuracy}]{\includegraphics[width=0.3\textwidth]{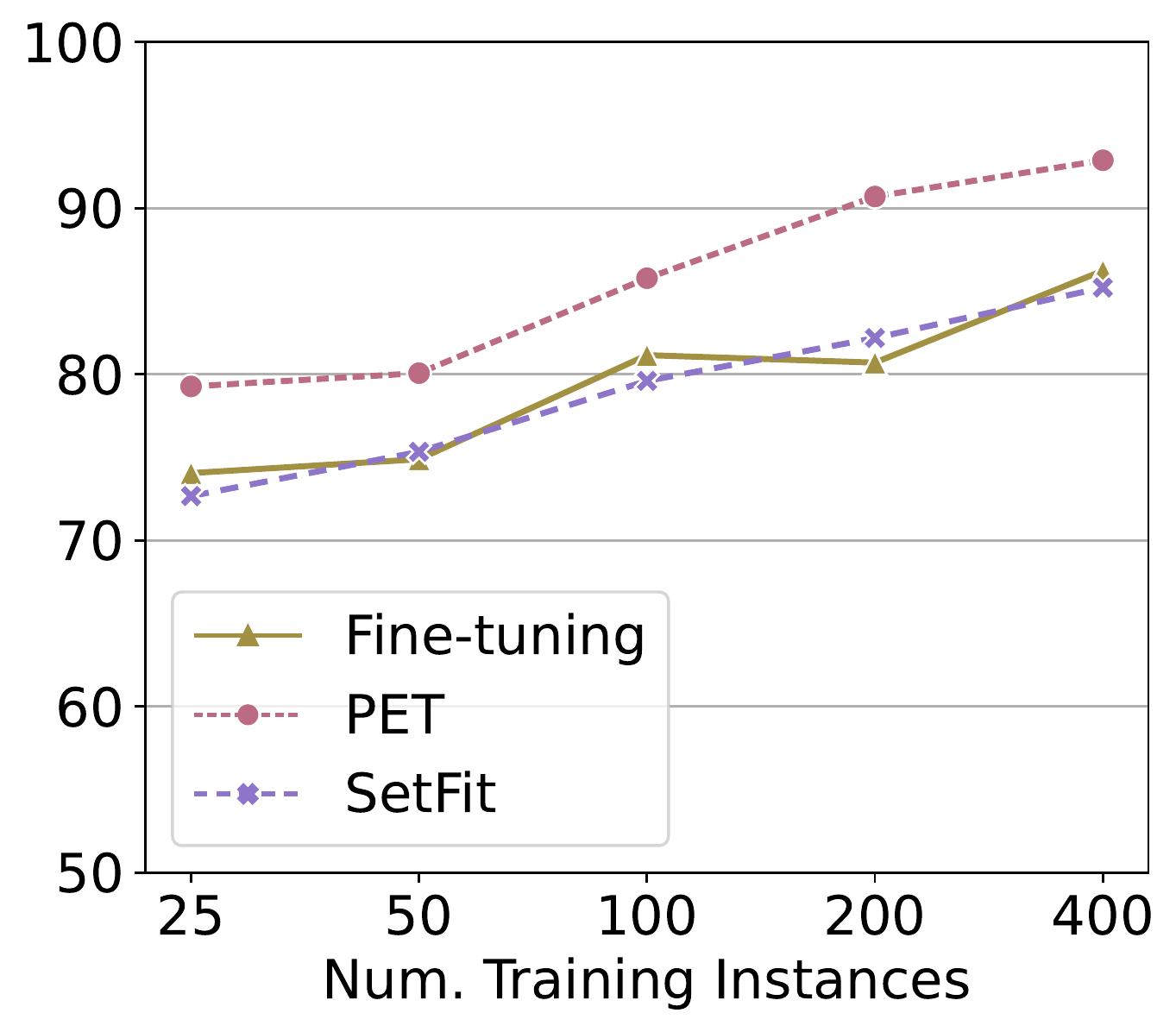}}
\\
\subfloat[\bertLarge{} (WF1)\label{fig:entailment-bert-large-uncased-weighted-avg-f1-score}]{\includegraphics[width=0.3\textwidth]{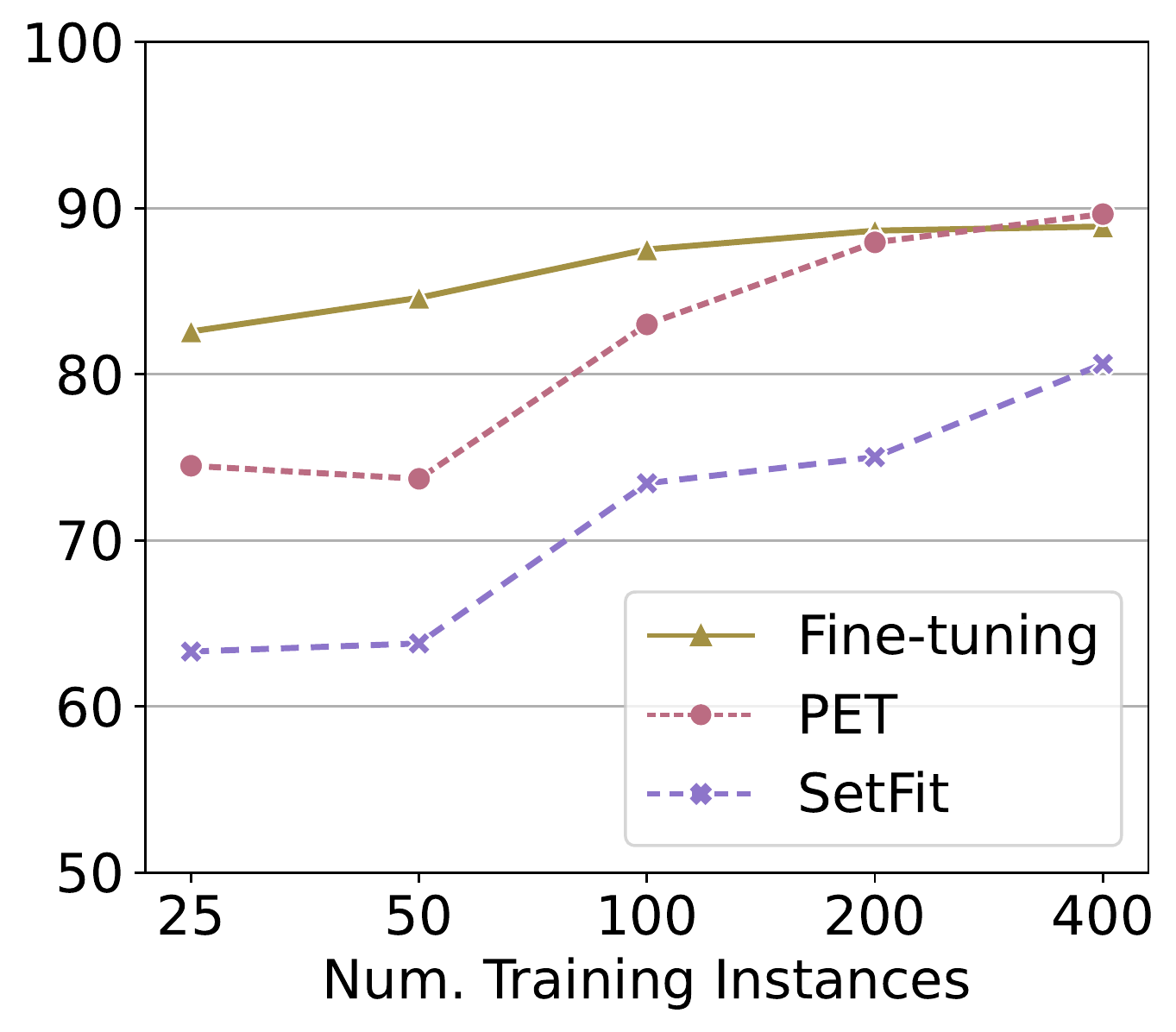}}
\subfloat[\robertaLarge{} (WF1)\label{fig:entailment-roberta-large-weighted-avg-f1-score}]{\includegraphics[width=0.3\textwidth]{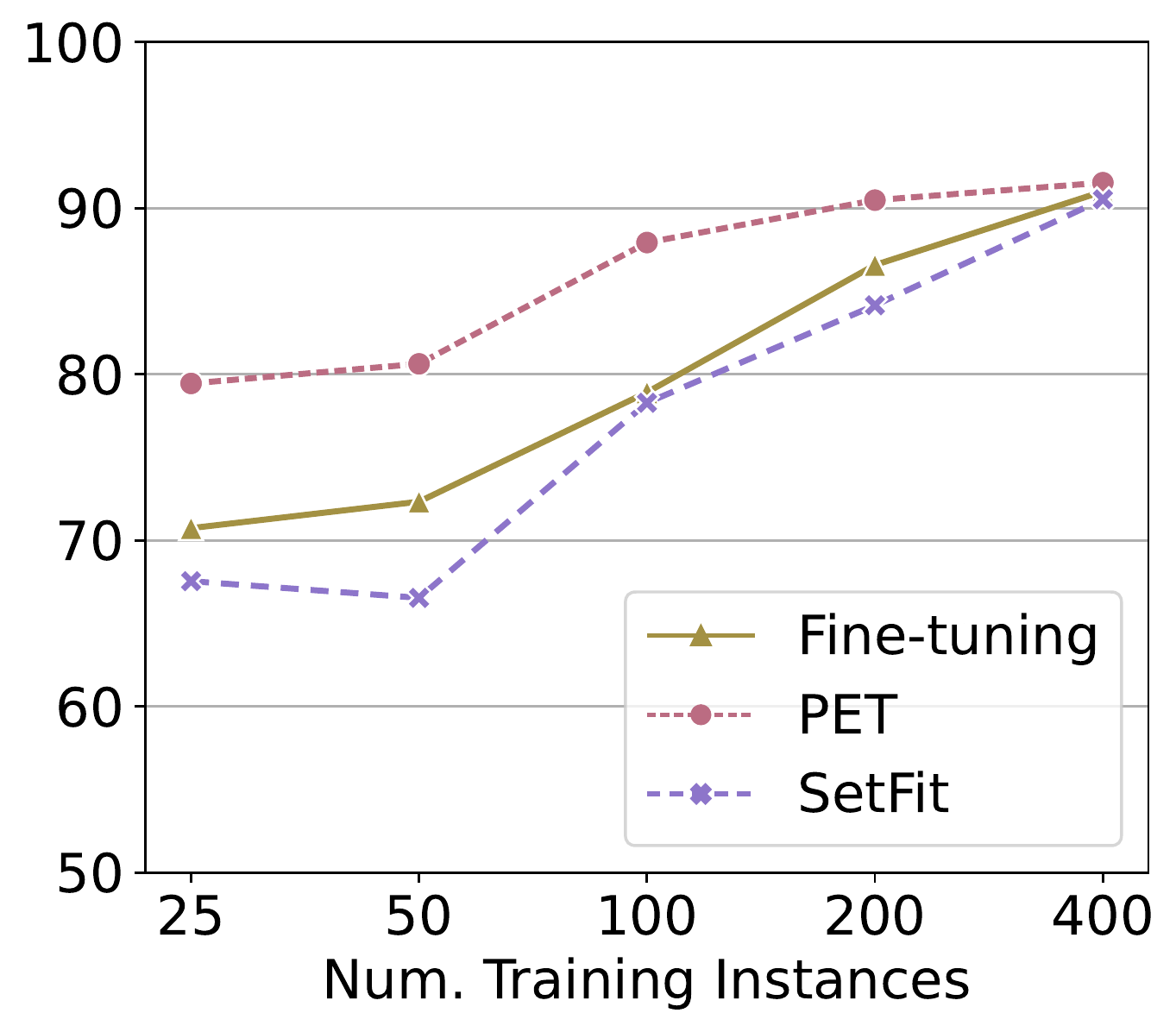}}
\subfloat[\debertaLarge{} (WF1)\label{fig:entailment-deberta-v3-large-weighted-avg-f1-score}]{\includegraphics[width=0.3\textwidth]{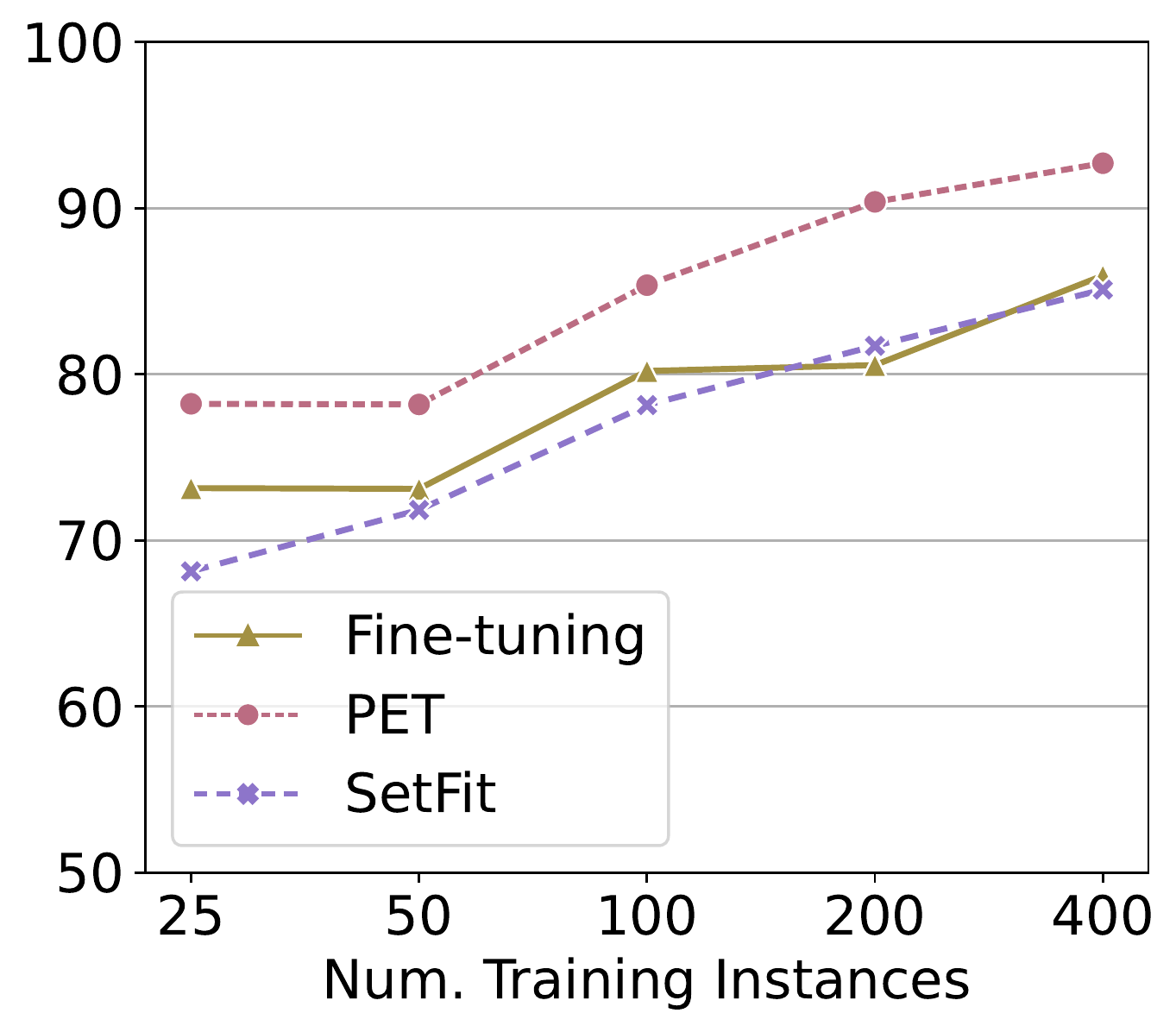}}
\\
\subfloat[\bertLarge{} (MF1)\label{fig:entailment-bert-large-uncased-macro-avg-f1-score}]{\includegraphics[width=0.3\textwidth]{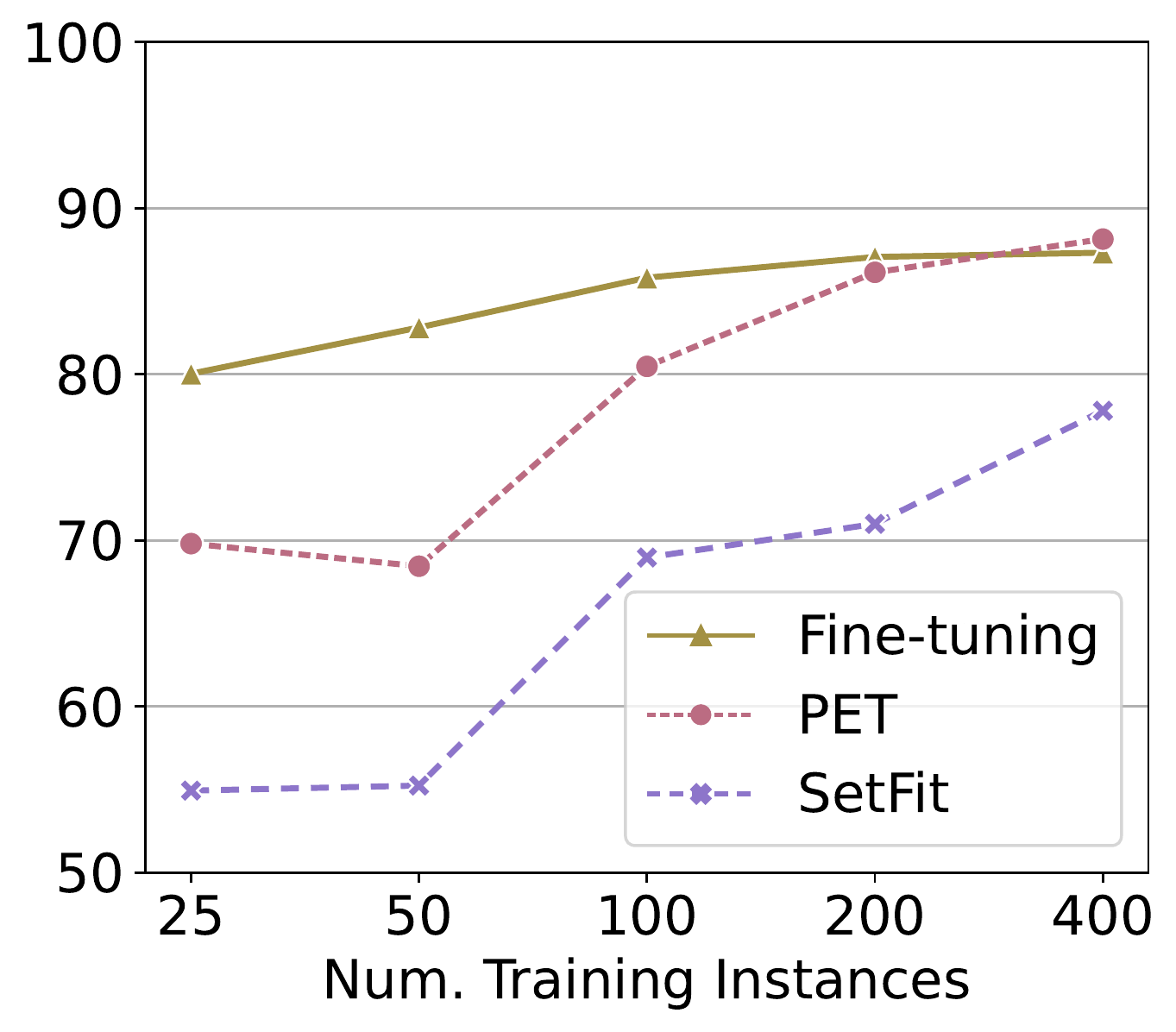}}
\subfloat[\robertaLarge{} (MF1)\label{fig:entailment-roberta-large-macro-avg-f1-score}]{\includegraphics[width=0.3\textwidth]{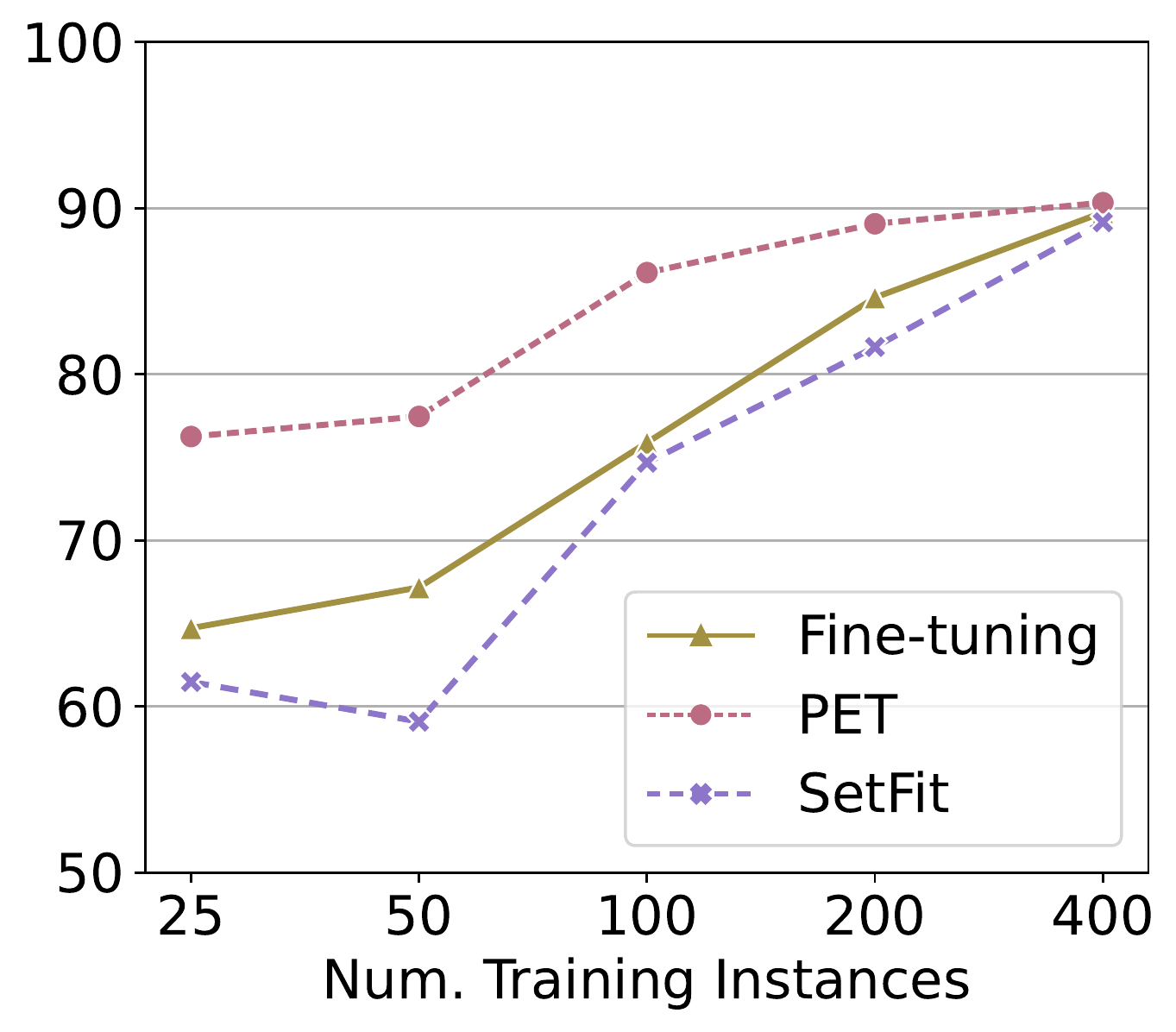}}
\subfloat[\debertaLarge{} (MF1)\label{fig:entailment-deberta-v3-large-macro-avg-f1-score}]{\includegraphics[width=0.3\textwidth]{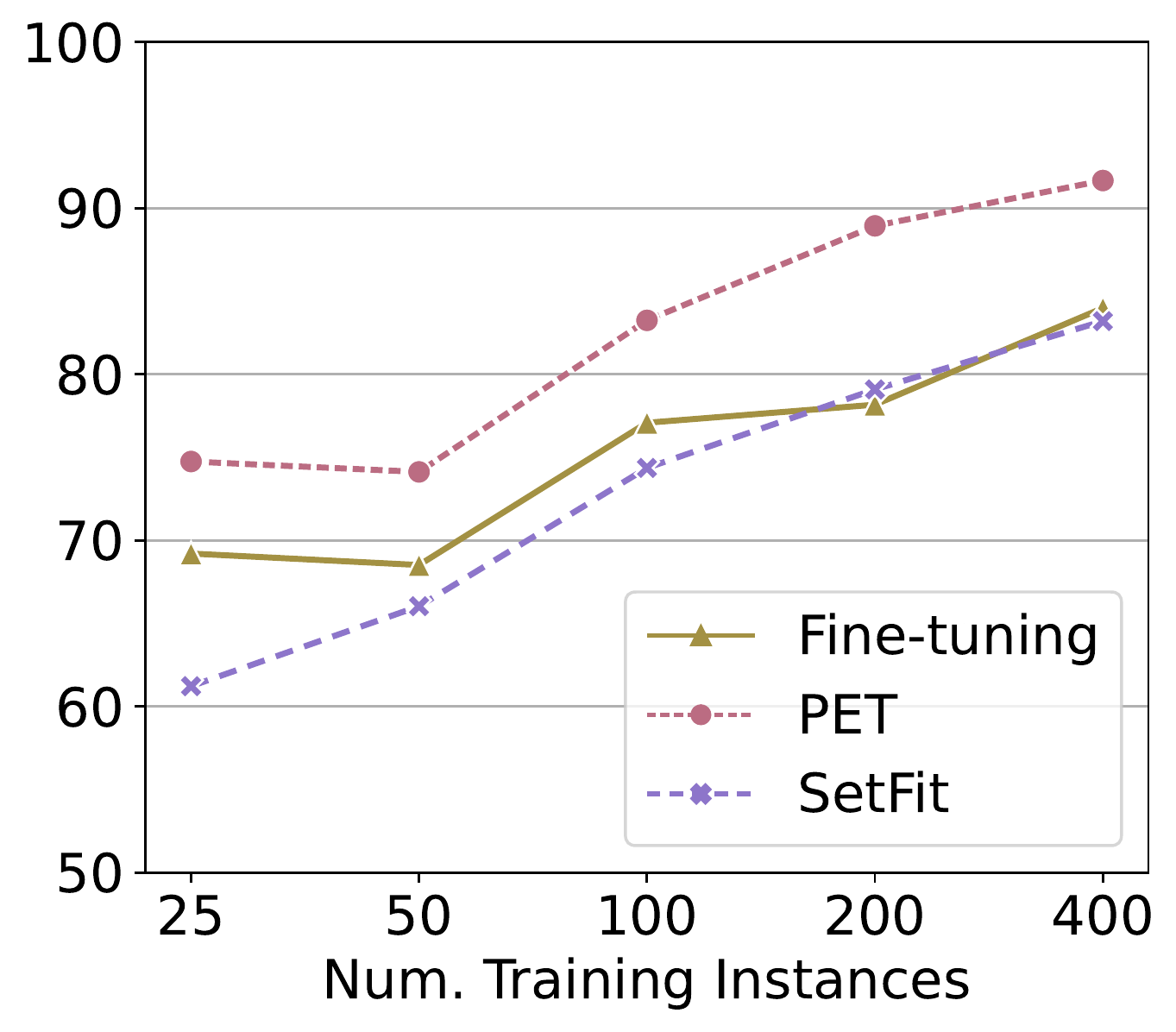}}
\\
\caption{Bugzilla bug dependency results.}
\begin{tablenotes}
\centering
\footnotesize
\item Acc: Accuracy
\item WF1: Weighted Average F1-Score
\item MF1: Macro Average F1-Score
\end{tablenotes}
\label{fig:entailment-all-results}
\end{figure}

We next consider the bug dependency detection (entailment) dataset, outlined in Figure~\ref{fig:entailment-all-results}.
Notably, both the Bugzilla duplicate detection dataset and the dependency detection dataset are scraped from Mozilla's Bugzilla bug report repository, but we observe substantially different results between these two datasets. 
Once again, fine-tuning is the preferred approach when using \bertLarge{}, and PET is the superior approach when using the other model checkpoints.
Unlike for duplicate detection, we observe that PET with \robertaLarge{} and PET with \debertaLarge{} is competitive with classic fine-tuning using \bertLarge{}.
That is, across all training set sizes, PET either outperforms fine-tuning with \bertLarge{} or is within a few percentage points of it.
We note that, regardless of the chosen checkpoint, the best SetFit is able to do is to remain competitive with fine-tuning, but it often scores 10\% or more lower than PET when using the same checkpoint.
The difference between PET with \robertaLarge{} and \debertaLarge{} is subtle across most training set sizes, but from \bertLarge{} to \robertaLarge{}, the performance improvements are considerably greater, with a gain of about 5\% points in accuracy for 25, 50, and 100 training instances.

\begin{figure}[!htb]
\centering
\subfloat[\bertLarge{} (Acc)\label{fig:stack-overflow-bert-large-uncased-accuracy}]{\includegraphics[width=0.3\textwidth]{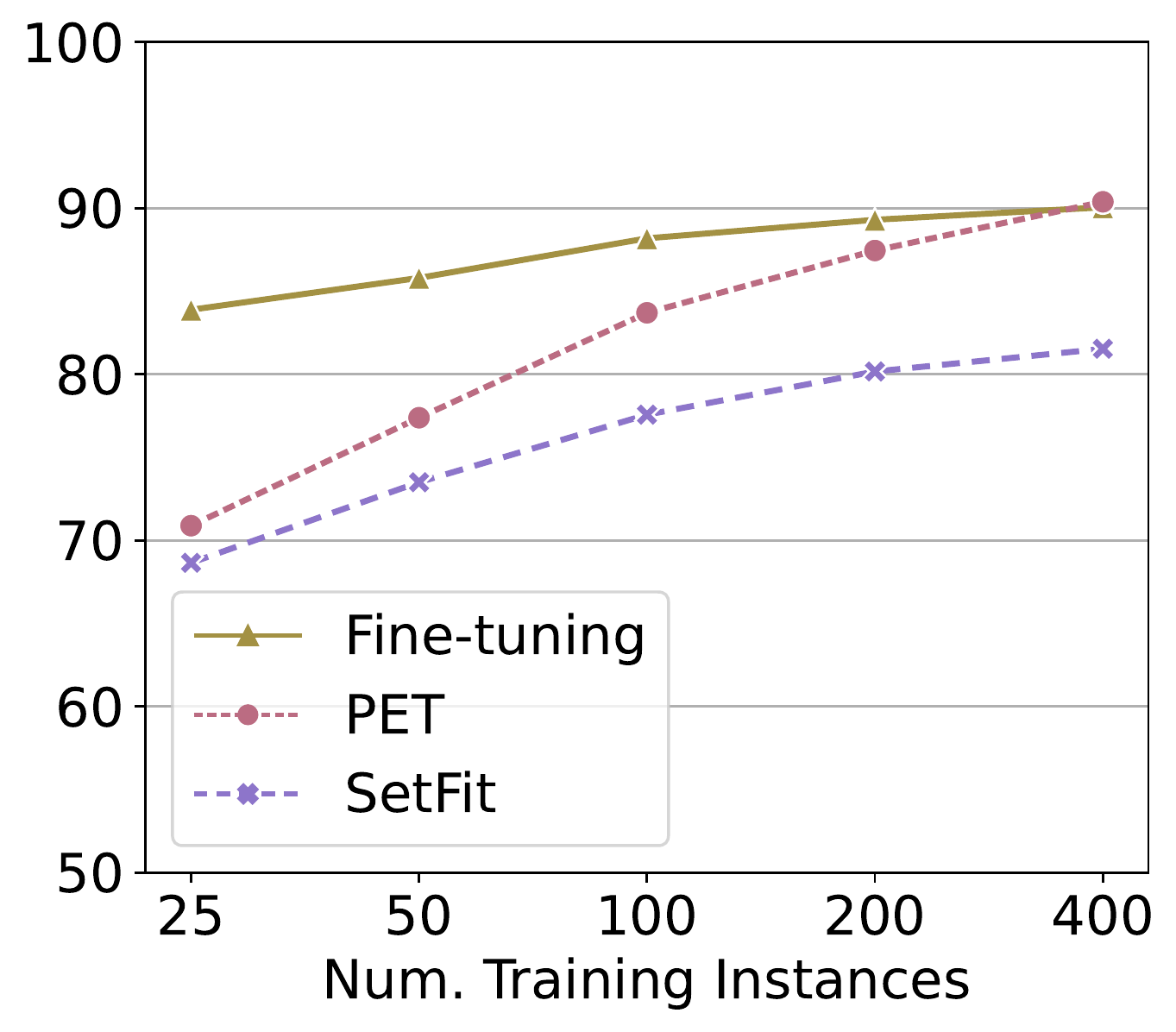}}
\subfloat[\robertaLarge{} (Acc)\label{fig:stack-overflow-roberta-large-accuracy}]{\includegraphics[width=0.3\textwidth]{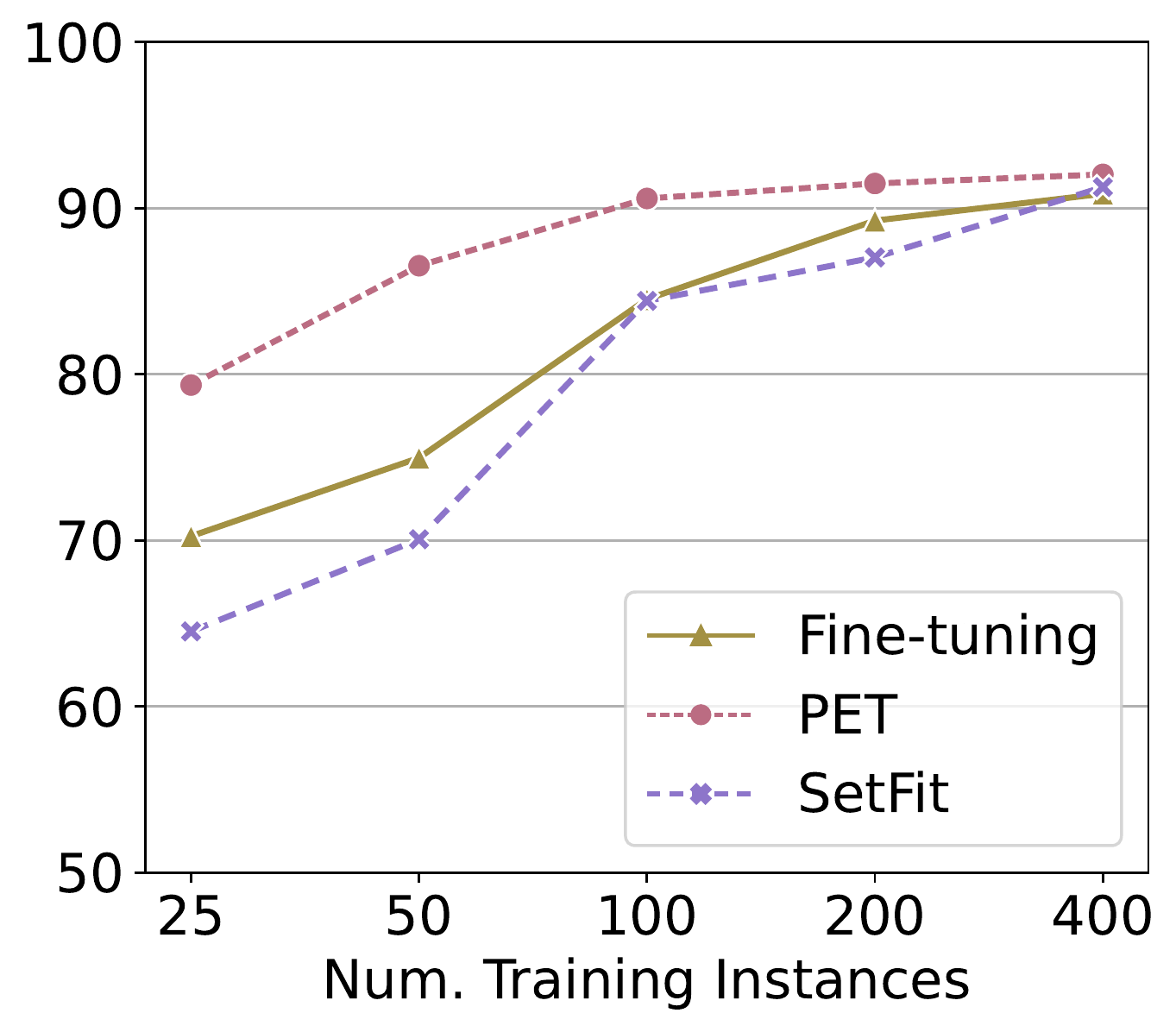}}
\subfloat[\debertaLarge{} (Acc)\label{fig:stack-overflow-deberta-v3-large-accuracy}]{\includegraphics[width=0.3\textwidth]{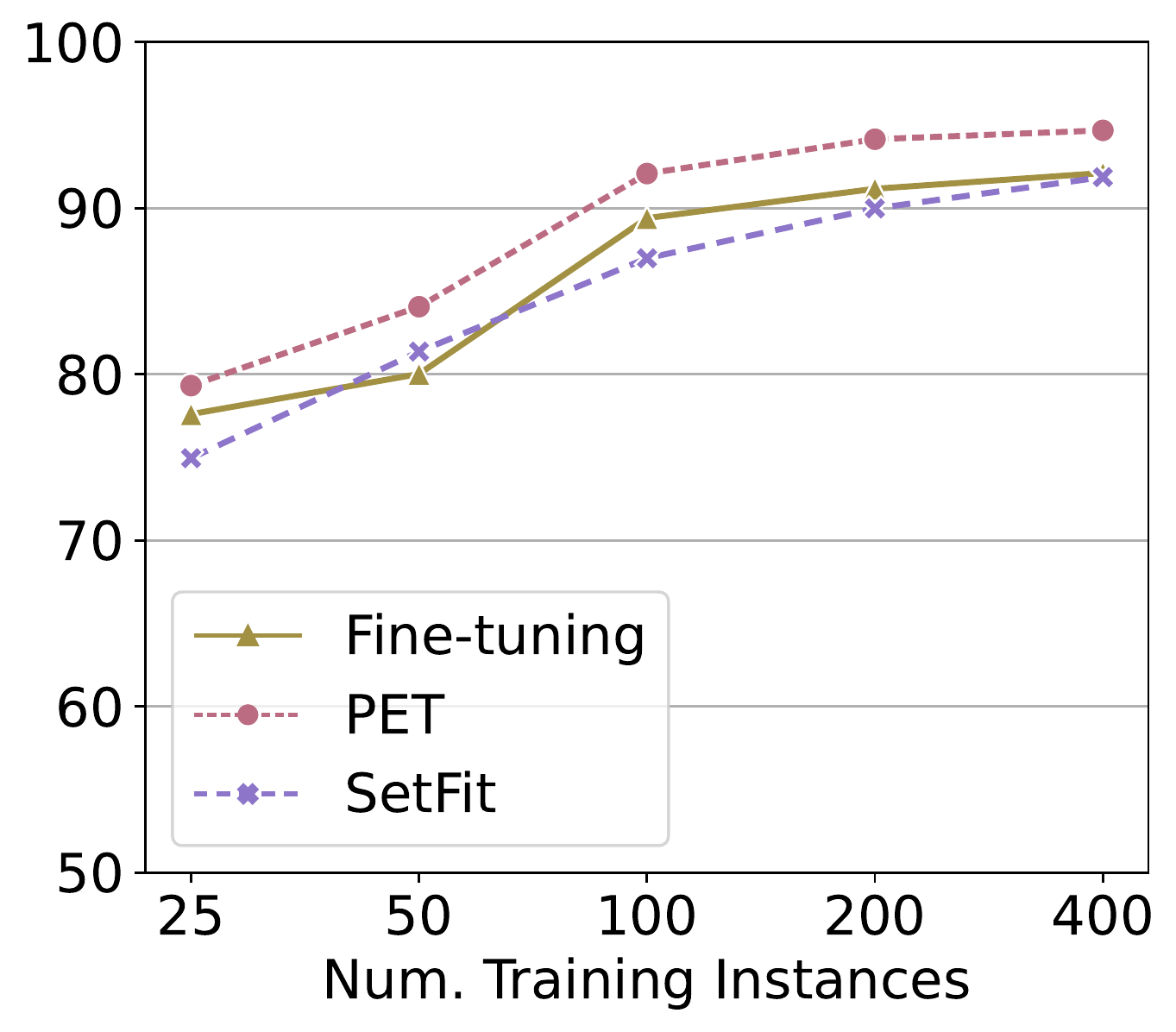}}
\\
\subfloat[\bertLarge{} (WF1)\label{fig:stack-overflow-bert-large-uncased-weighted-avg-f1-score}]{\includegraphics[width=0.3\textwidth]{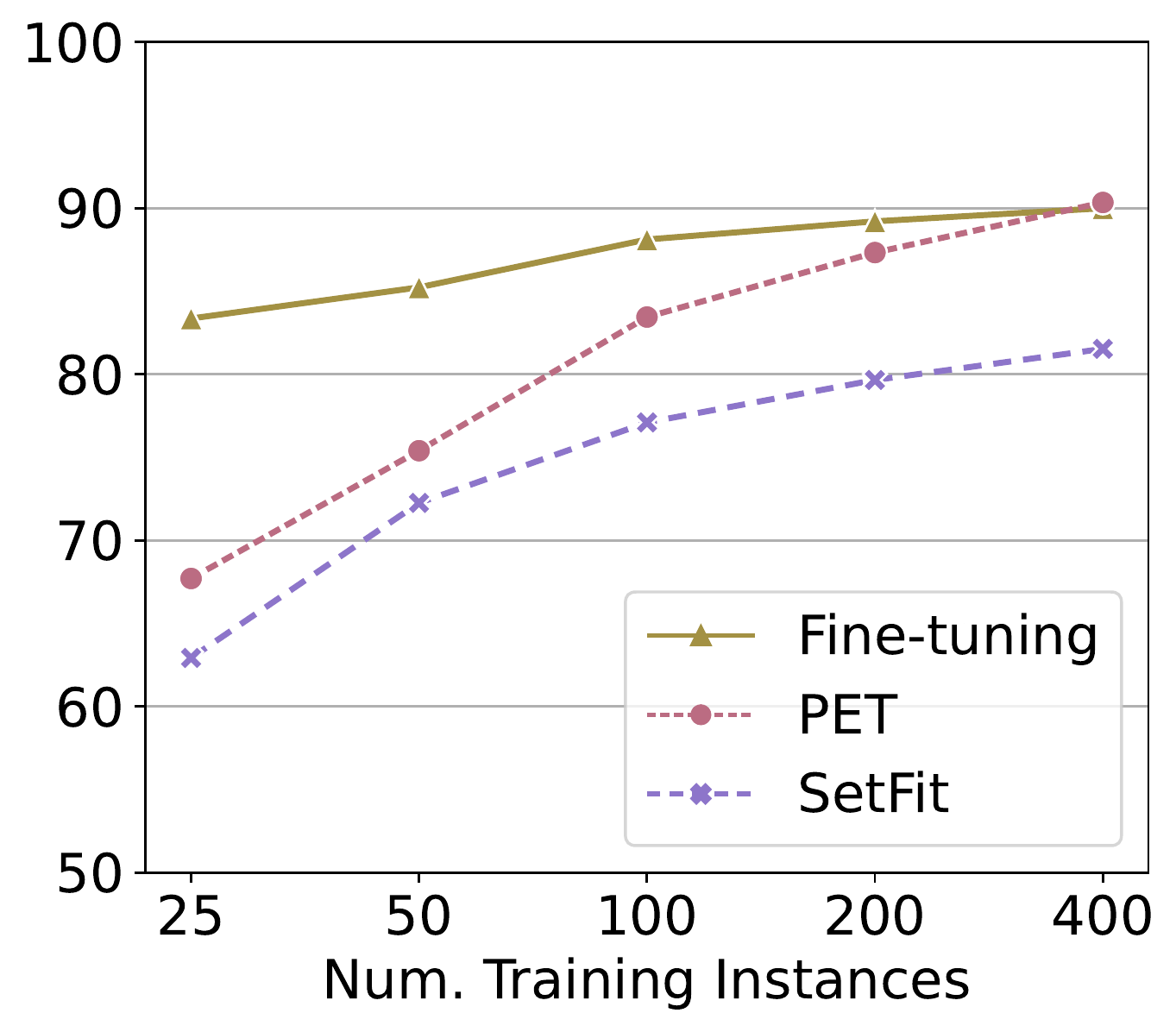}}
\subfloat[\robertaLarge{} (WF1)\label{fig:stack-overflow-roberta-large-weighted-avg-f1-score}]{\includegraphics[width=0.3\textwidth]{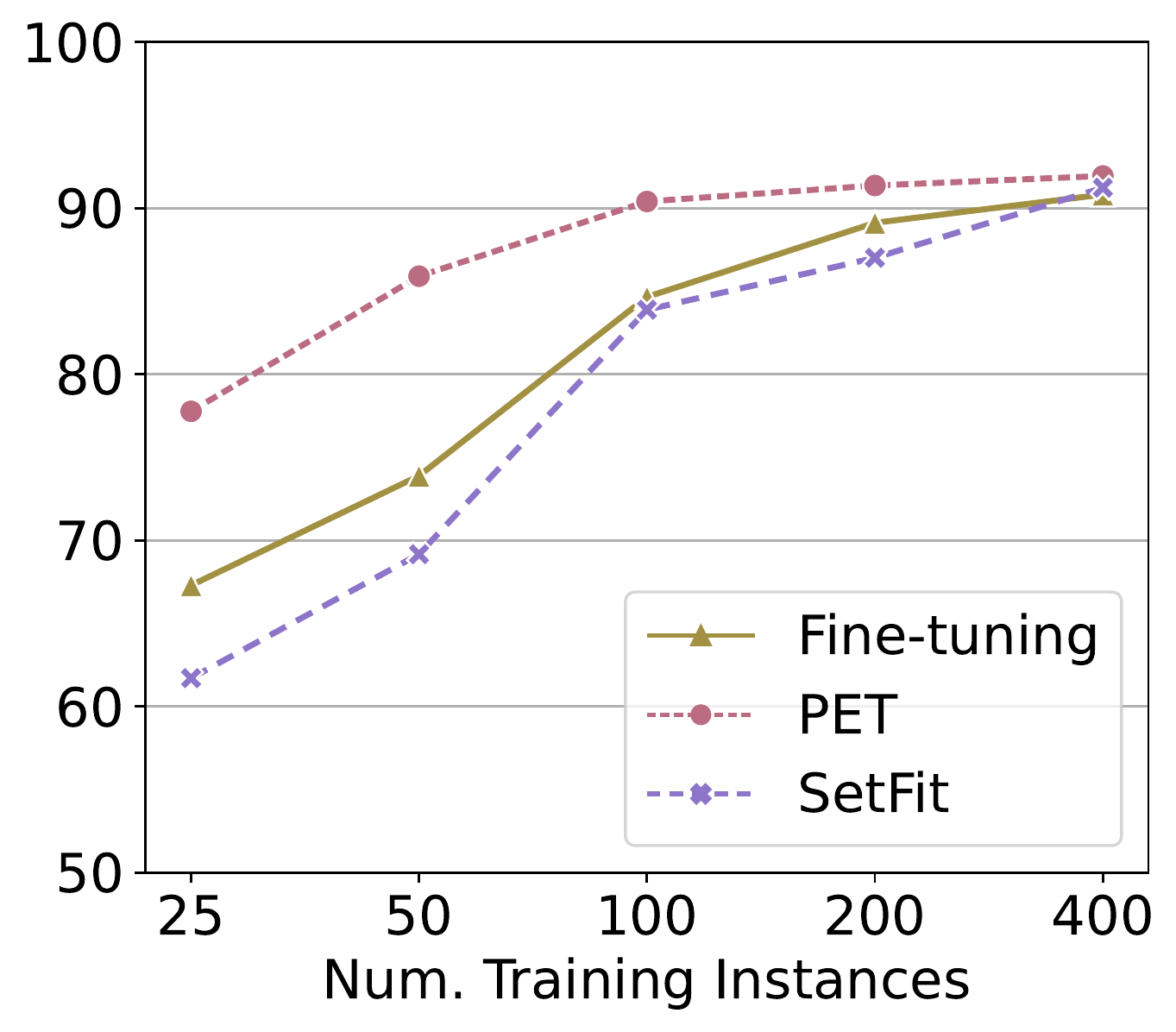}}
\subfloat[\debertaLarge{} (WF1)\label{fig:stack-overflow-deberta-v3-large-weighted-avg-f1-score}]{\includegraphics[width=0.3\textwidth]{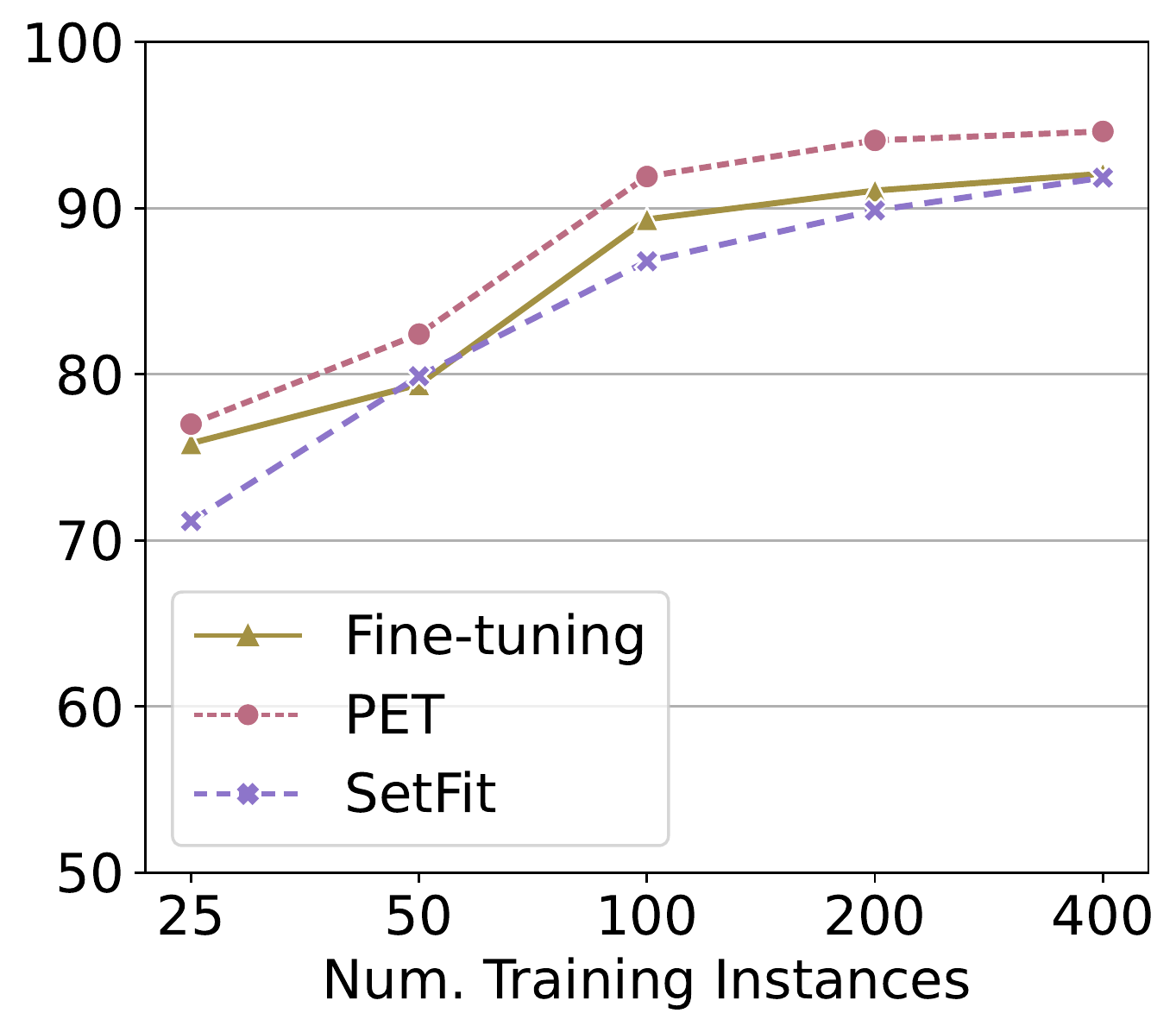}}
\\
\subfloat[\bertLarge{} (MF1)\label{fig:stack-overflow-bert-large-uncased-macro-avg-f1-score}]{\includegraphics[width=0.3\textwidth]{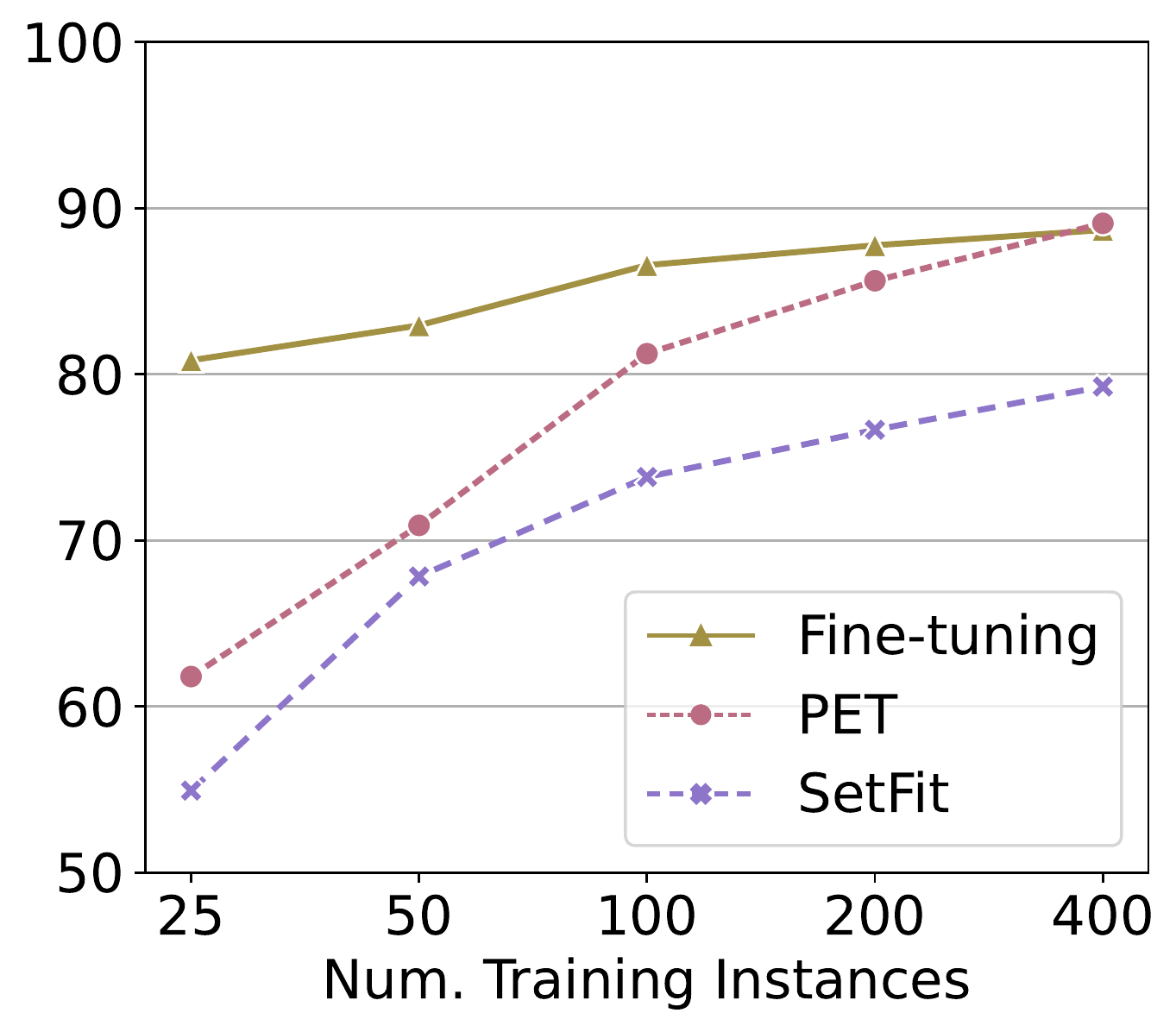}}
\subfloat[\robertaLarge{} (MF1)\label{fig:stack-overflow-roberta-large-macro-avg-f1-score}]{\includegraphics[width=0.3\textwidth]{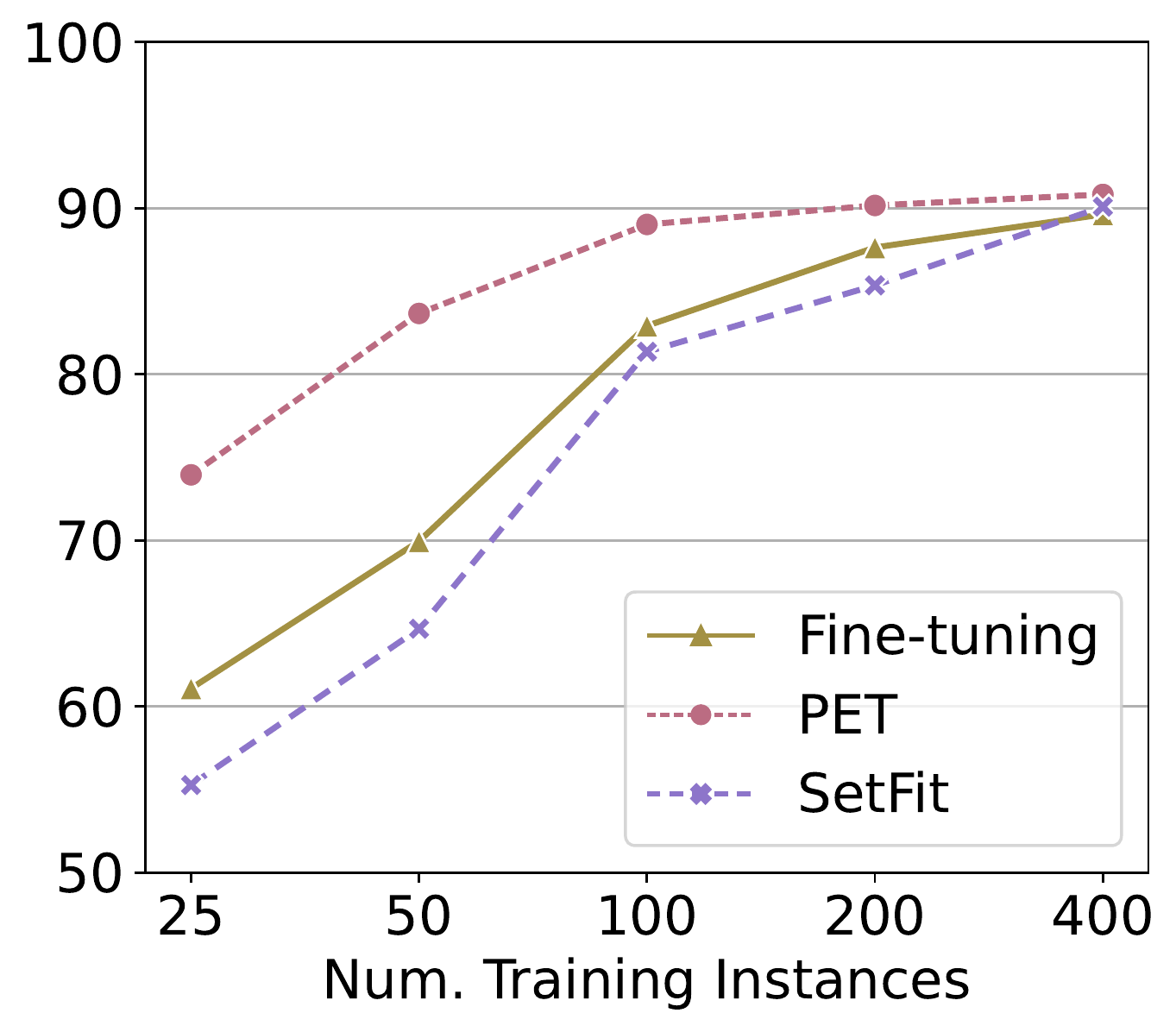}}
\subfloat[\debertaLarge{} (MF1)\label{fig:stack-overflow-deberta-v3-large-macro-avg-f1-score}]{\includegraphics[width=0.3\textwidth]{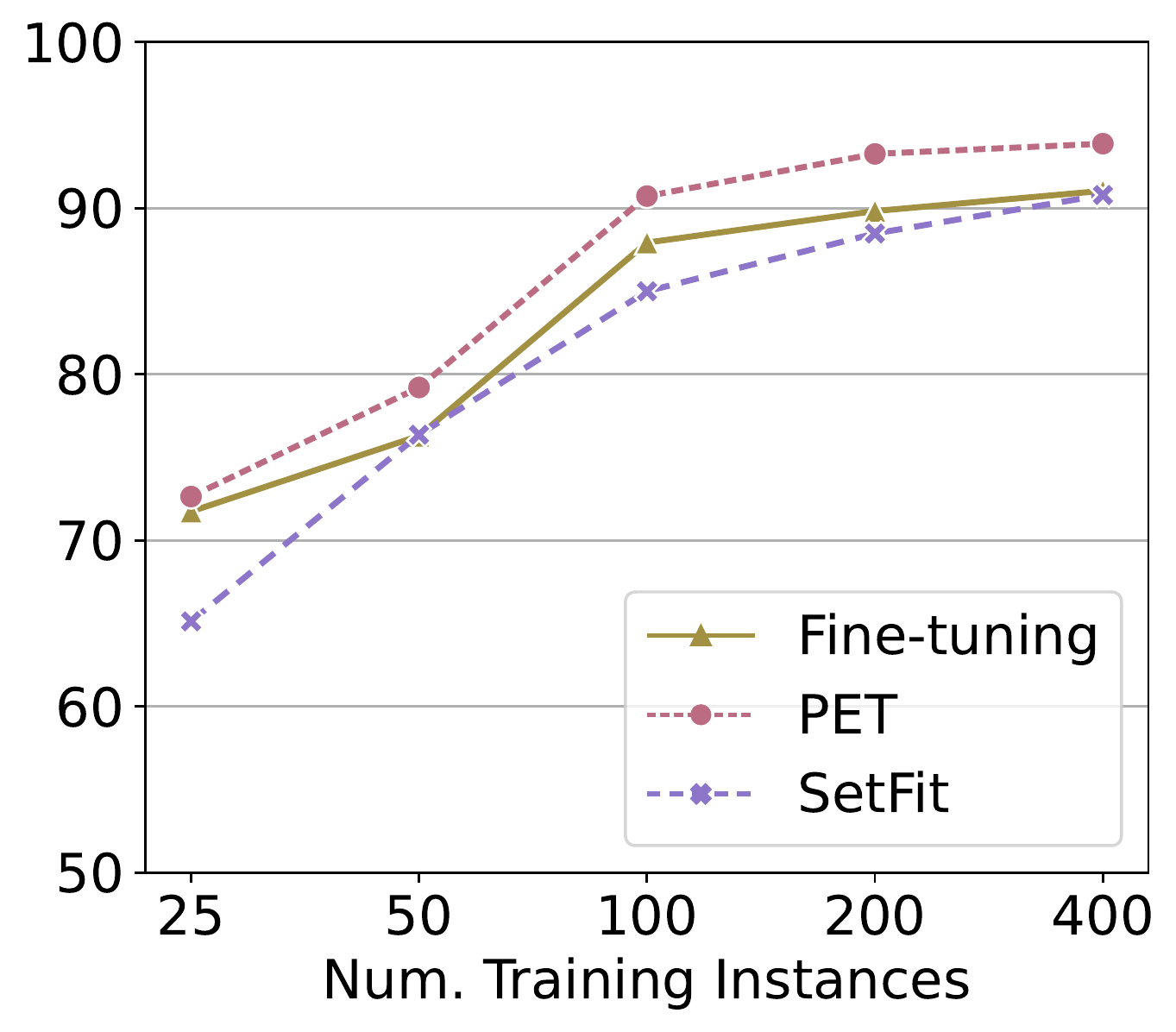}}
\\
\caption{Stack Overflow duplicate detection results.}
\begin{tablenotes}
\centering
\footnotesize
\item Acc: Accuracy
\item WF1: Weighted Average F1-Score
\item MF1: Macro Average F1-Score
\end{tablenotes}
\label{fig:stack-overflow-all-results}
\end{figure}

Figure~\ref{fig:stack-overflow-all-results} gives the results for the Stack Overflow duplicate detection dataset.
We observe that for 25 training instances, PET with \robertaLarge{} is just a few percentage points in accuracy behind fine-tuning with \bertLarge{}.
However, when we consider the macro average F1-score for these two results, the gap is more than 5\%.
As early as 50 labeled examples the PET algorithm with \robertaLarge{} becomes the leader for this task, overtaken by PET with \debertaLarge{} for the largest dataset sizes.
As with our earlier results, we do not observe SetFit to be competitive with the other strategies when using the same checkpoint.
However, for smaller training sets, SetFit with \debertaLarge{} outperforms fine-tuning with \robertaLarge{} and PET with \bertLarge{}.
Unfortunately, when compared to PET and fine-tuning with the same underlying language model, SetFit still consistently offers the weakest performance.

\begin{figure}[!htb]
\centering
\subfloat[\bertLarge{} (Acc)\label{fig:conflict-bert-large-uncased-accuracy}]{\includegraphics[width=0.3\textwidth]{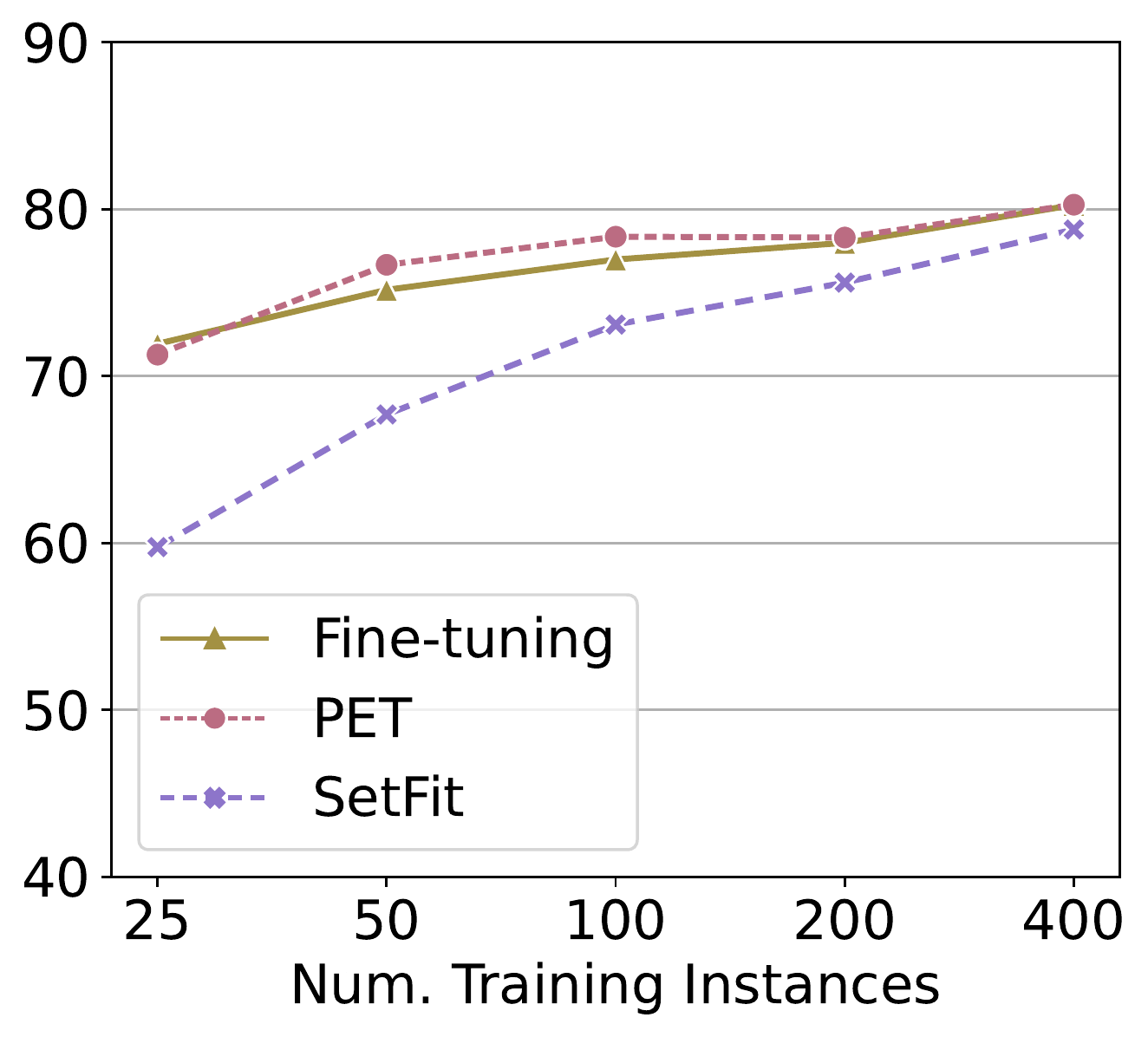}}
\subfloat[\robertaLarge{} (Acc)\label{fig:conflict-roberta-large-accuracy}]{\includegraphics[width=0.3\textwidth]{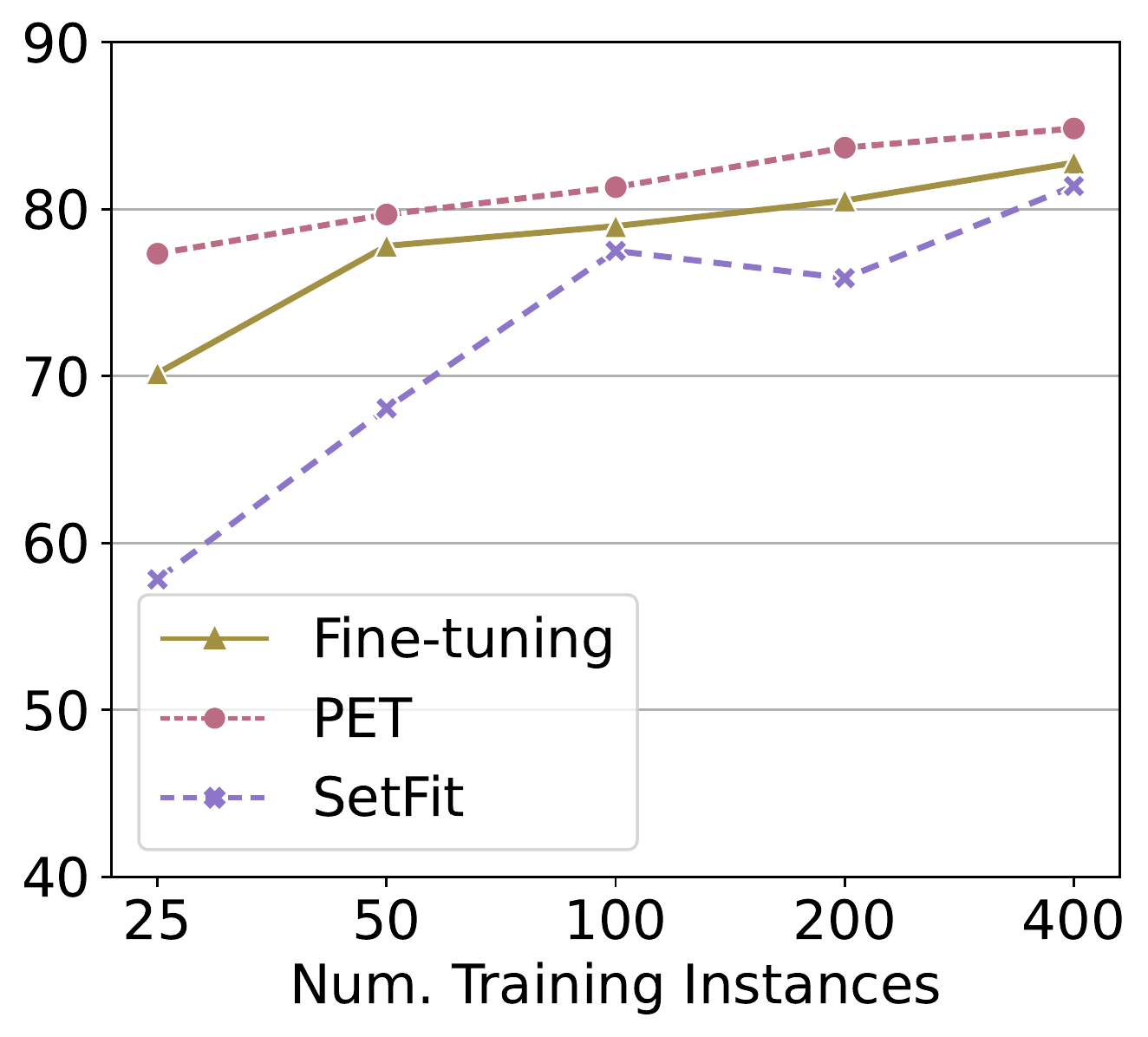}}
\subfloat[\debertaLarge{} (Acc)\label{fig:conflict-deberta-v3-large-accuracy}]{\includegraphics[width=0.3\textwidth]{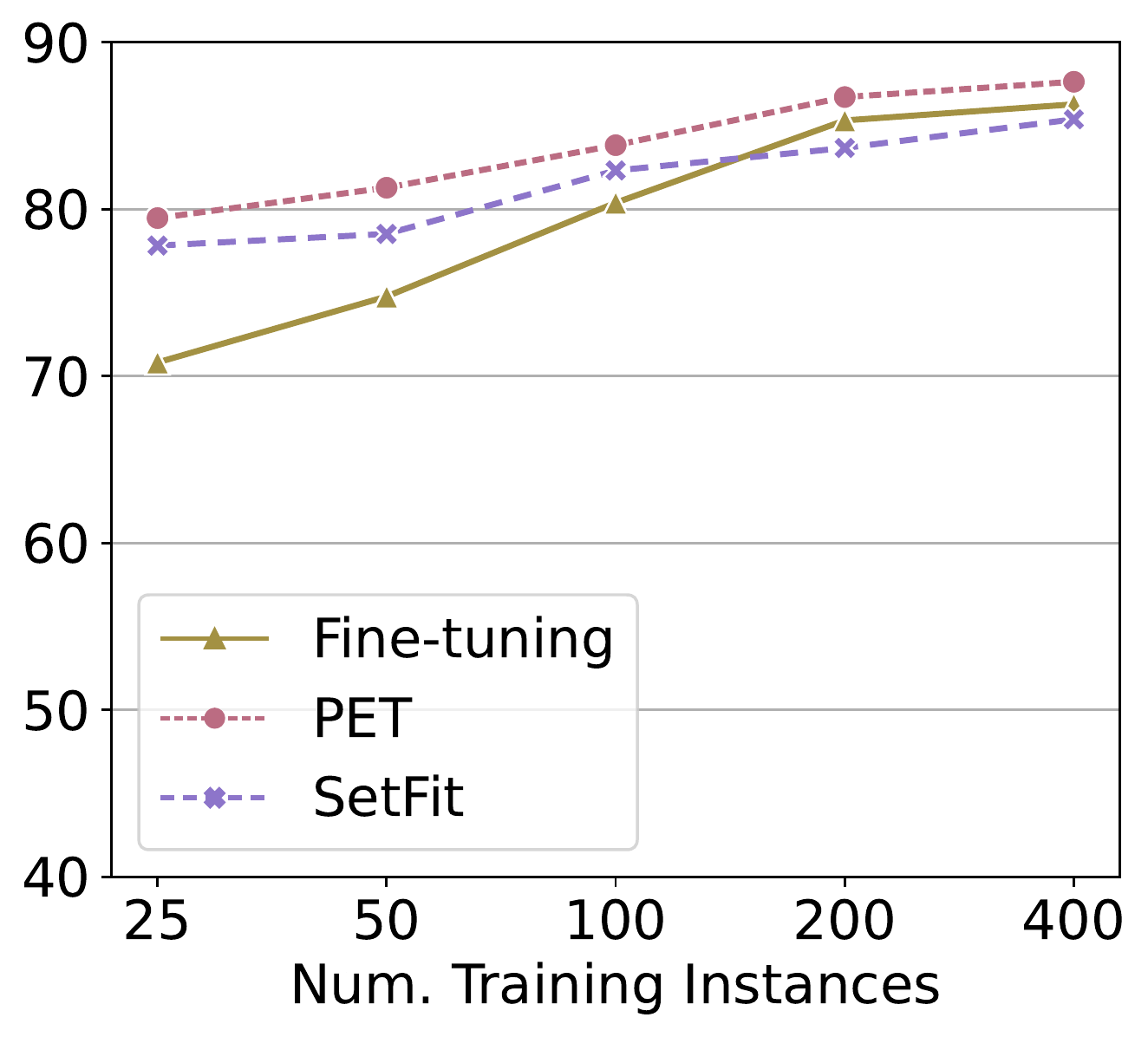}}
\\
\subfloat[\bertLarge{} (WF1)\label{fig:conflict-bert-large-uncased-weighted-avg-f1-score}]{\includegraphics[width=0.3\textwidth]{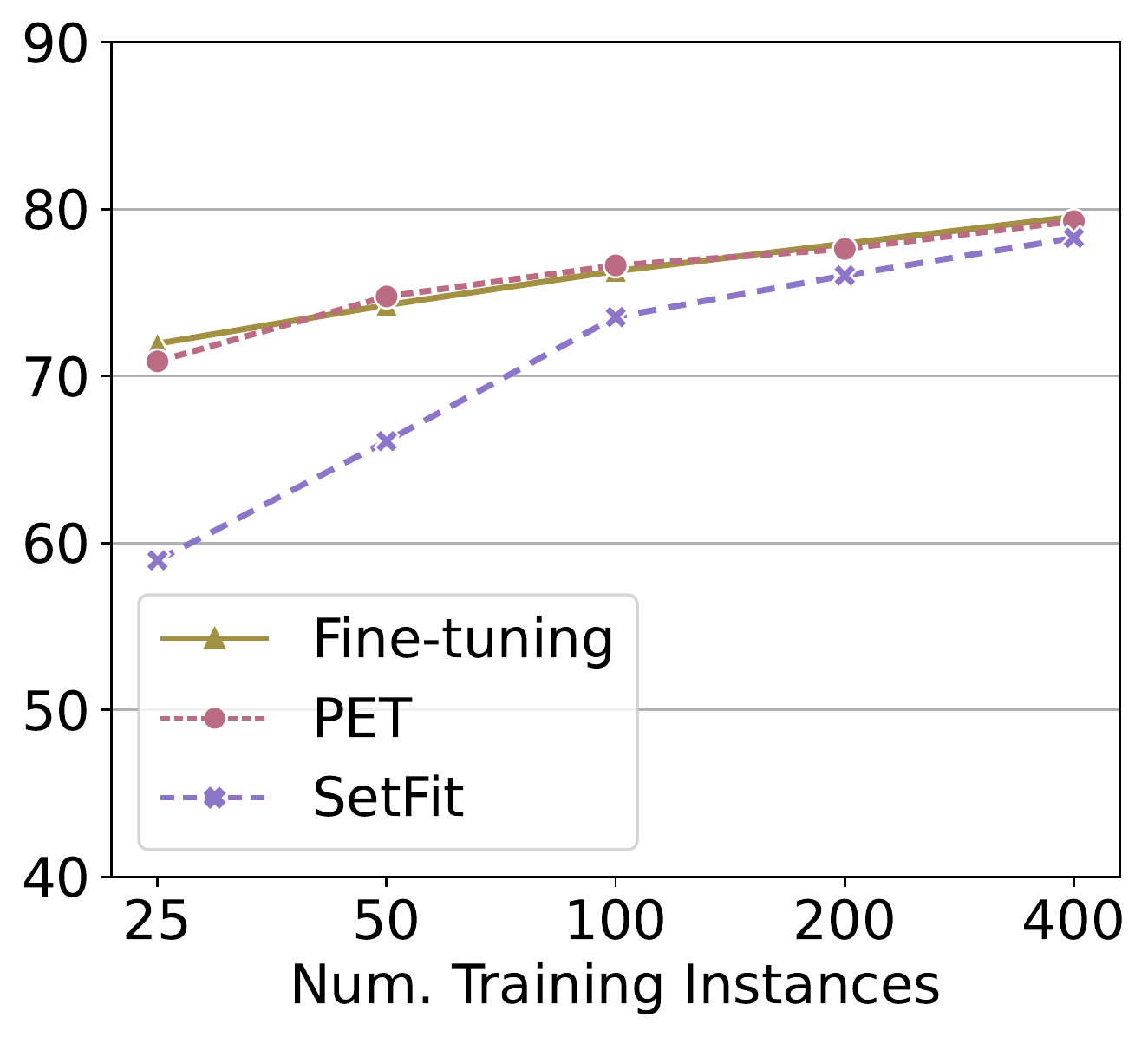}}
\subfloat[\robertaLarge{} (WF1)\label{fig:conflict-roberta-large-weighted-avg-f1-score}]{\includegraphics[width=0.3\textwidth]{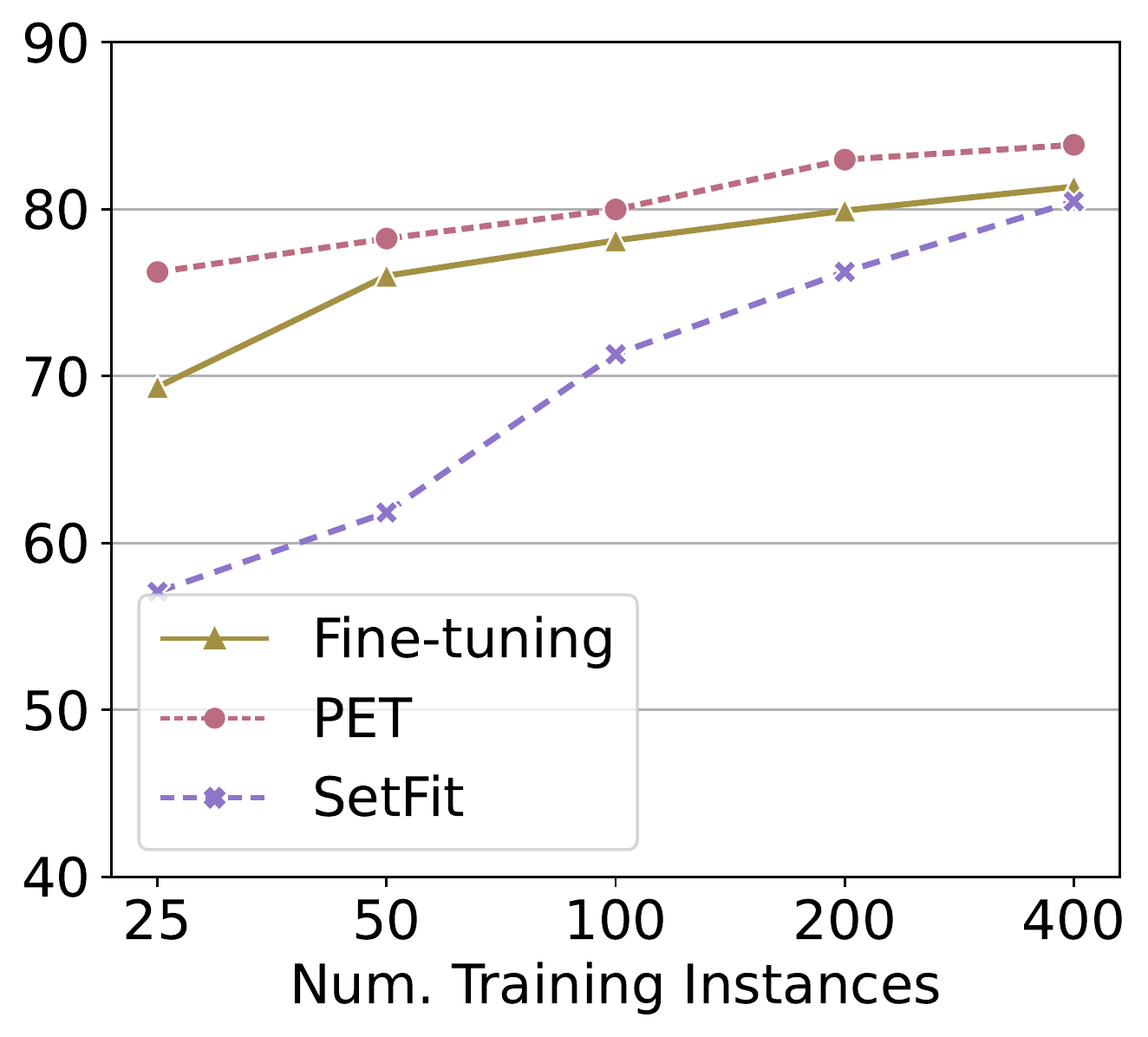}}
\subfloat[\debertaLarge{} (WF1)\label{fig:conflict-deberta-v3-large-weighted-avg-f1-score}]{\includegraphics[width=0.3\textwidth]{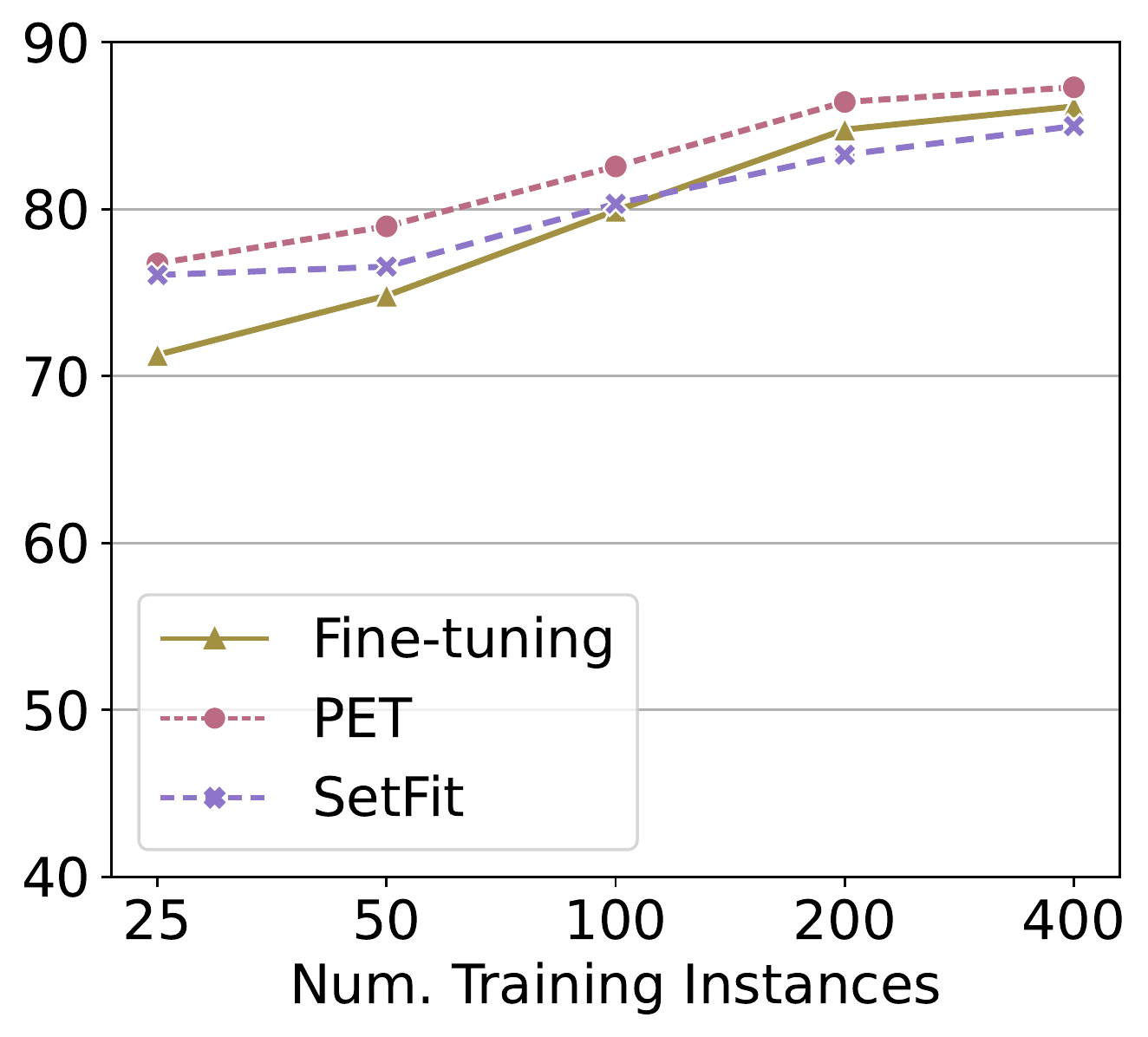}}
\\
\subfloat[\bertLarge{} (MF1)\label{fig:conflict-bert-large-uncased-macro-avg-f1-score}]{\includegraphics[width=0.3\textwidth]{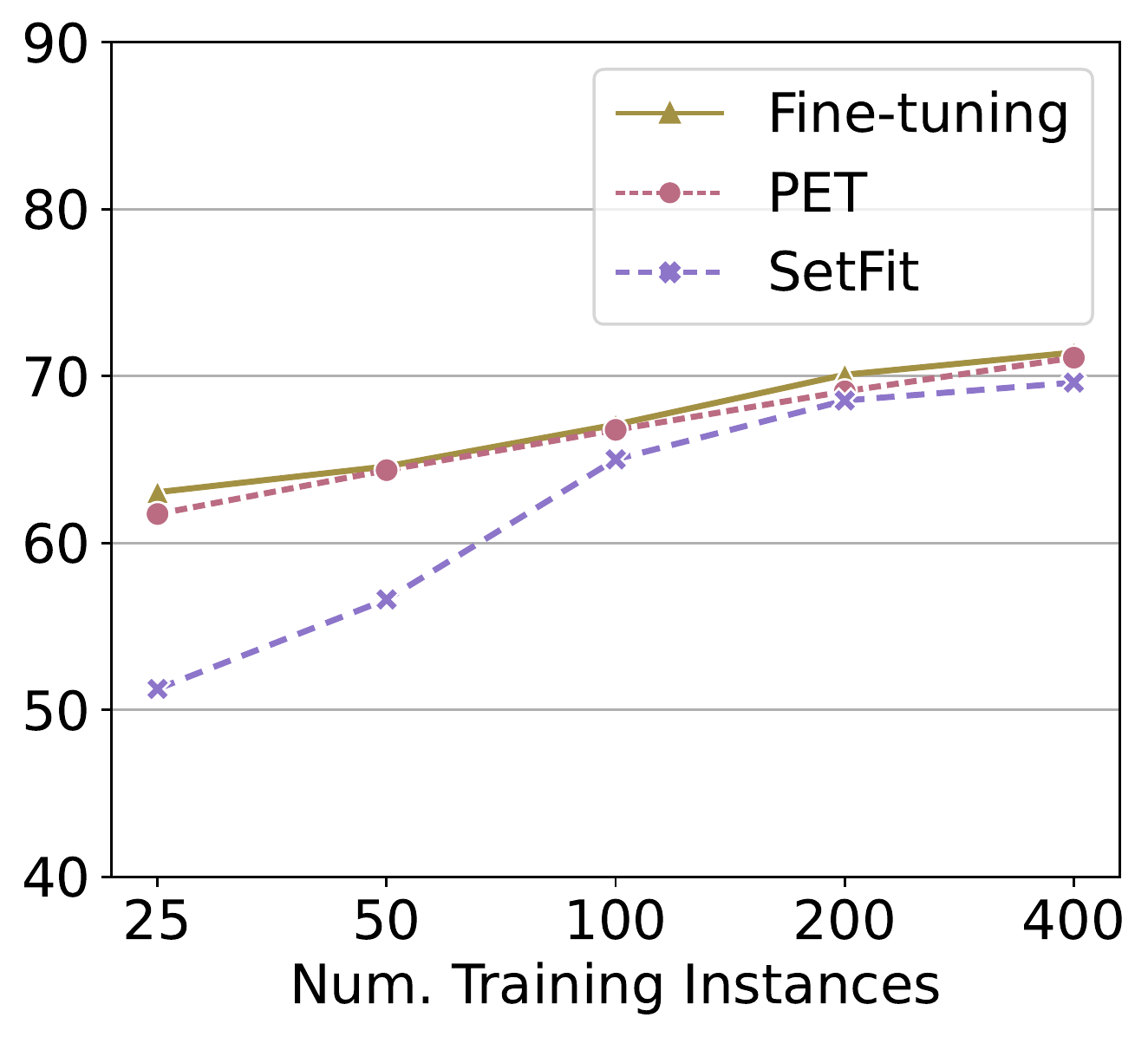}}
\subfloat[\robertaLarge{} (MF1)\label{fig:conflict-roberta-large-macro-avg-f1-score}]{\includegraphics[width=0.3\textwidth]{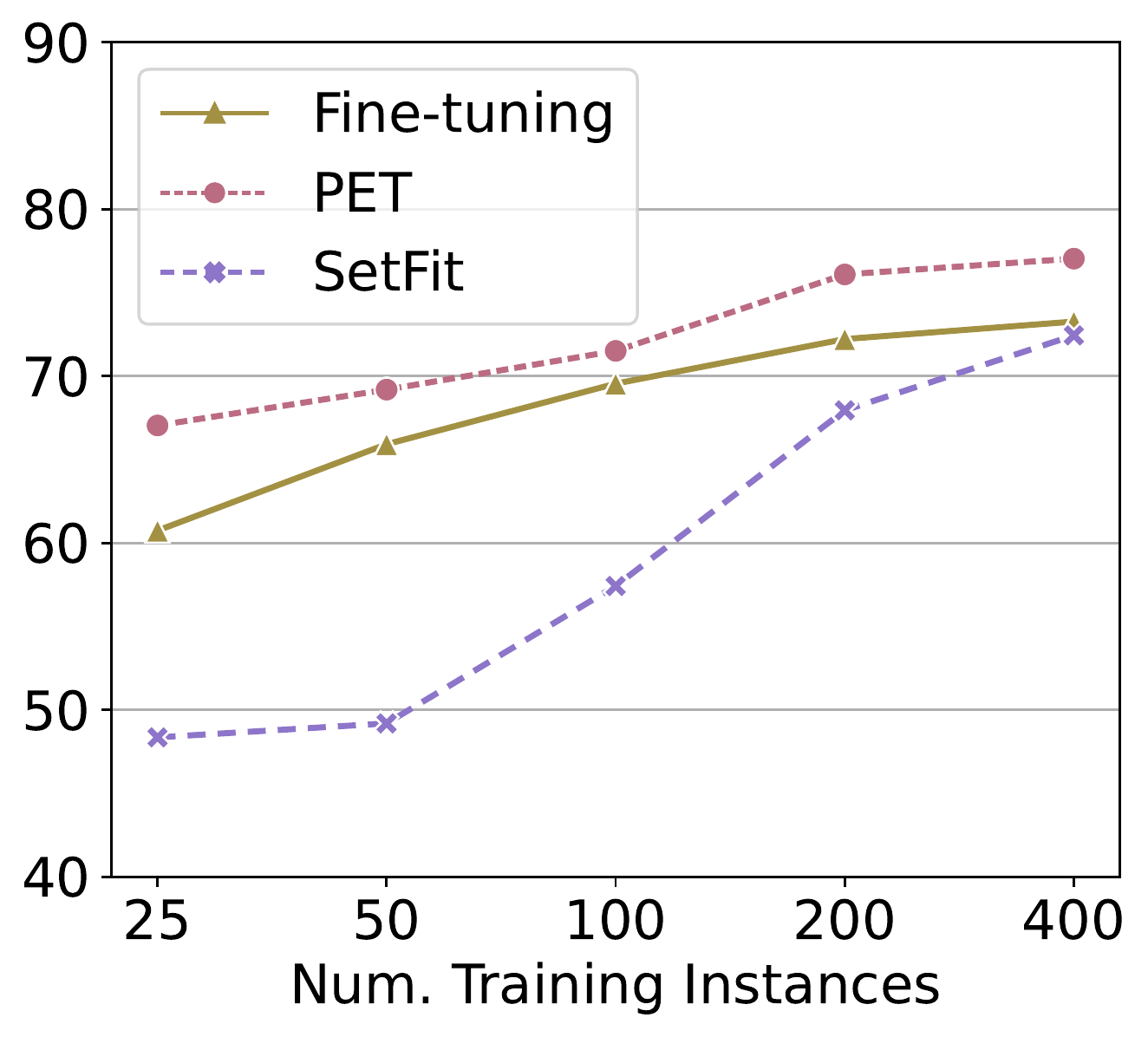}}
\subfloat[\debertaLarge{} (MF1)\label{fig:conflict-deberta-v3-large-macro-avg-f1-score}]{\includegraphics[width=0.3\textwidth]{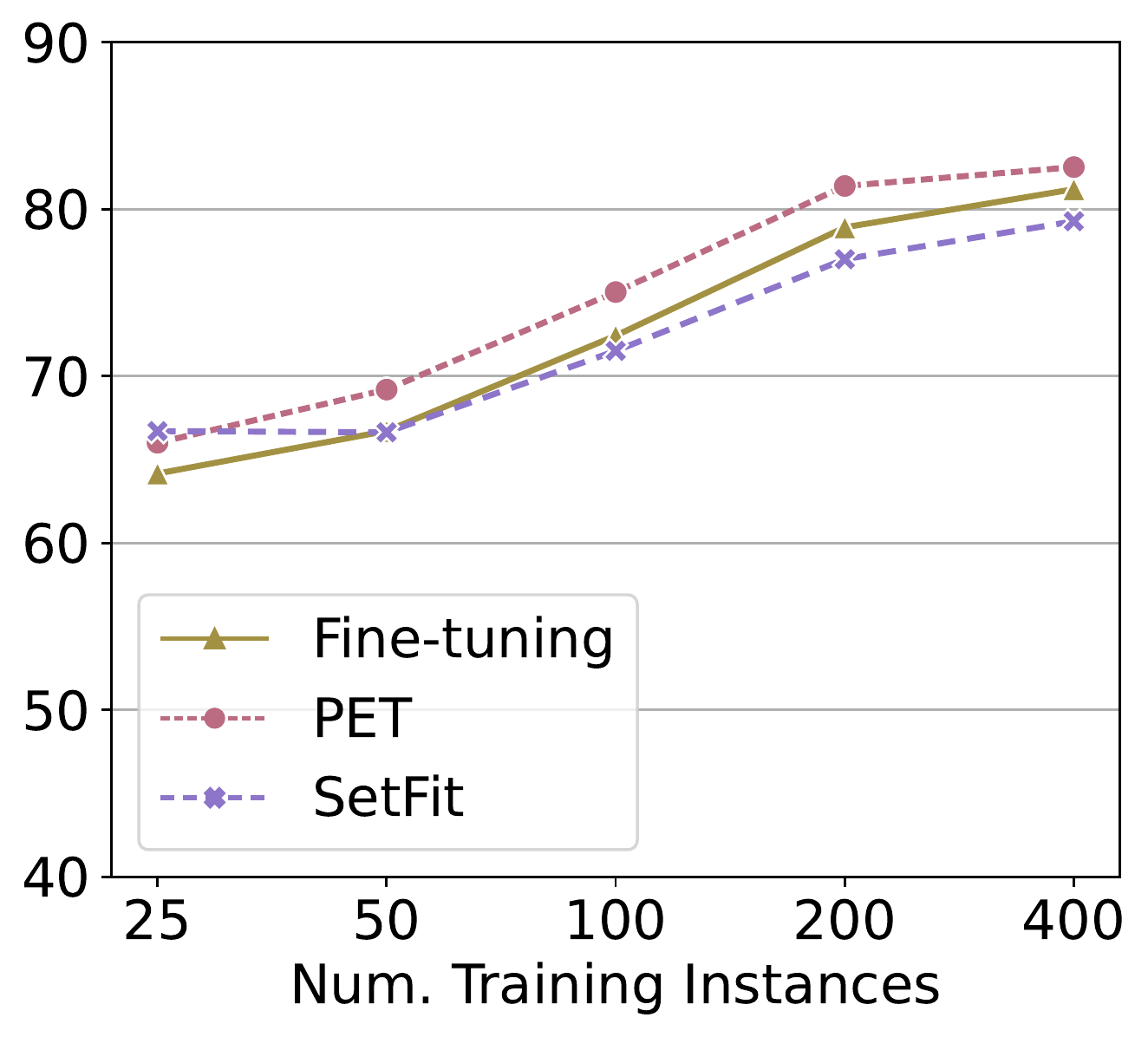}}
\\
\caption{SRS conflict detection results.}
\begin{tablenotes}
\centering
\footnotesize
\item Acc: Accuracy
\item WF1: Weighted Average F1-Score
\item MF1: Macro Average F1-Score
\end{tablenotes}
\label{fig:conflict-all-results}
\end{figure}

Finally, we consider the conflict detection task.
Figure~\ref{fig:conflict-all-results} outlines the performance of each model on the conflict detection task.
This is the only software engineering task that we consider which shows PET with \bertLarge{} performing similarly to fine-tuning with \bertLarge{}.
For this problem, we observe that PET with a \debertaLarge{}-base tends to be the best performing model across all metrics.
We also find that SetFit with \debertaLarge{} outperforms fine-tuning for smaller dataset sizes, and also generally outperforms or matches the performance of the approaches using the other model checkpoints.
The results for SetFit are also much more stable here, with mostly a clear pattern of increasing performance for increasing training set size.

\section{Discussion and Conclusions}\label{sec:conclusion}
This work focuses on few-shot learning algorithms and their applications in various software engineering tasks.
We provide a detailed overview of the current literature, and investigate fine-tuning, as well as the SetFit~\citep{tunstall2022efficient} and pattern-exploiting training (PET)~\citep{schick2021exploiting} methods.
In our analysis, we compare each of these approaches on three transformer model checkpoints: \bertLarge{} \citep{devlin2019bert}, \robertaLarge{} \citep{liu2019roberta}, and \debertaLarge{} \citep{he2021deberta}.
In the current literature, it is often the case that a single model is selected as the basis for few-shot learning approaches.
Accordingly, our investigation provides an overview of the relative performance of each model for each checkpoint.
We also measure the robustness of the hyperparameters used for each model, as in a few-shot setting changing additional hyperparameters beyond the model checkpoint is not possible without validation data, which we assume to be unavailable.
Our results also investigate the performance of each algorithm for increasing training set size in order to determine the upside of training models with more than just a few training data, but substantially less data than those would be considered as a full-sized data set.
Finally, we evaluate each model on a variety of tasks and, in the case of software engineering, we derive these tasks from different problems and data sources to provide a more detailed overview of few-shot learning performance.

In conducting our research, we attempted to curate diverse data sets and problem instances to ensure a thorough investigation of few-shot learning in the software engineering domain.
However, our research is not without its limitations.
Firstly, we note that when selecting the few-shot learning approaches to experiment with, we consider performance on the RAFT leaderboard~\citep{alex2021raft} to determine the state-of-the-art.
A major limitation of the RAFT leaderboard is that it does not contain any data sets which use sentence pairs, meaning that the tasks explored in this study are different from the data used to evaluate models on the RAFT leaderboard.
As in the context of few-shot learning there is no development set, the chosen algorithm, checkpoint, and hyperparameters must be chosen based on an educated guess, and leaderboards provide a fair way to inform such decisions.
Accordingly, we believe that the next iteration of the RAFT leaderboard can benefit from including sentence pair tasks, as this will allow for better informed decisions in the context of model selection in those cases.
Secondly, our model checkpoints are all of a similar architecture.
Specifically, \robertaLarge{} and \debertaLarge{} are effectively incremental improvements over \bertLarge{}.
While our results showed that \bertLarge{} was still able to outperform these models for some tasks in a few-shot setting, experimenting with a more diverse array of model checkpoints may reveal different relationships between few-shot learning approaches than observed in this work.

We summarize the key takeaways into a list of recommendations.
With the understanding that they should be considered together with the limitations of our study, we provide the following recommendations:
\begin{itemize}
    \item \textbf{Label as many examples as you can.} 
    While it is easy to fixate on which method performs the best when only e.g., 25 examples are available, in practice one will almost always be concerned with maximizing model performance.
    When we consider a fixed approach and a fixed checkpoint, we observe that the difference in performance between 25 training instances and 400 training instances is frequently 20 absolute percentage points or more for all metrics across most of the problems considered.
    This represents a substantial improvement and still requires labeling less data than in a full-sized context.
    Accordingly, if you require the best performance, consider labeling hundreds of examples rather than tens.
    \item \textbf{Select the best method for the task you are working on.}
    While leaderboards like \citet{alex2021raft}'s RAFT provide an ``Overall'' score to help compare the performance of few-shot learning approaches, the results here show that one cannot consider overall performance alone to draw conclusions.
    You should look for model performance on datasets similar to the one you are considering.
    This means look for datasets covering a similar task, with a similar number of labels, which consider the same sentence classification task (i.e., if you are using sentence pairs, look for a similar dataset using sentence pairs).
    \item \textbf{Do not select a model based on performance in a data rich environment.}
    On the SuperGLUE leaderboard,\footnote{\url{https://super.gluebenchmark.com/leaderboard}} \debertaLarge{} is a top-performer.
    It also performed well on the full-sized datasets.
    However, we frequently observed that it performed poorly when only 25 training instances were available.
    As we increased the training set size, we did observe that it became a top performer both for fine-tuning and \robertaLarge{}, but it is clear that choosing a model based on full-size dataset performance will not yield the best results.
\end{itemize}

There are many opportunities for future research from this work.
Firstly, future research may investigate the use of ensemble methods with fine-tuning and SetFit.
In general, this may offer a more fair performance comparison with the PET algorithm, which is ensemble-based.
Secondly, we note that we considered only bug summaries and Stack Overflow question titles when handling sentence pair classification.
As a single-sentence summary/title is not always available for a given task, an investigation into the performance of few-shot learning approaches for large bodies of text would offer useful insight into this field.
Thirdly, as this work focuses on classifying sentence pairs, an investigation into single-sentence classification is necessary for software engineering applications.
Fourthly, future research may consider additional models pre-trained using NSP loss to evaluate its impact on sentence pair classification tasks.

\section*{Disclosure Statement}
The authors have no relevant financial or non-financial interests to disclose.

\section*{Data Availability Statement}
The datasets generated during and/or analysed during the current study are available from the corresponding author on reasonable request.

\bibliographystyle{elsarticle-harv}
\bibliography{Ref}

%
%
%
%

\appendix

\section{Alternative few-shot learning approaches}

PET is a highly popular algorithm in the literature.
As of today, the combined number of citations on Google Scholar\footnote{\url{https://scholar.google.ca}} for \citet{schick2021exploiting} and \citet{schick2021size} exceeds 1000.
Naturally, many authors have investigated methods for improving upon the performance of the PET algorithm.
We discuss a few of these variants in the literature here and why we chose to use PET over these variants.

Firstly, we review Iterative PET (iPET), which was introduced in the original PET paper \citep{schick2021exploiting}.
iPET involves training generations of PET models on increasingly larger training sets.
Specifically, the training data is used to train a set of masked language models $\setModel$.
Then, for each model $\instanceModel\in \setModel$, a new training set is created by combining the original labeled training data with a unique portion of the unlabeled data.
This unlabeled data is given soft labels by a subset of the models in $\setModel$.
Accordingly, after the first generation, each model $\instanceModel$ is trained on a different set of training data.
At each iteration, the amount of unlabeled data used is increased and this process is repeated for several iterations.
Finally, the complete unlabeled dataset is given soft labels, and a sequence classifier is trained on the union of the labeled training set and the softly-labeled unlabeled dataset as in traditional PET. 
\citet{schick2021exploiting} reported that iPET performed better than PET in all cases.
The empirical results in the literature often show that iPET only offers a small performance improvement at the cost of significantly increased training time. Accordingly, we do not investigate its use here.

\citet{tam2021adapet} propose ADAPET, an extension of PET which does not require access to unlabeled data.
The foundation for ADAPET is a redesigned loss function which consists of two components: \textit{decoupled label loss} and \textit{label-conditioned MLM loss}.
The decoupled label loss modifies how the algorithm treats tokens that do not belong to the subset of tokens in the verbalizer.
Specifically, in the PET algorithm, the score for a given label $\petScore(\instanceLabel\|\petPattern_\instanceModel(\petEx), \instanceModel)$ is calculated by considering only those tokens represented in the verbalizer.
That is, only tokens $\instanceVocabulary$ for which there exists a label $\instanceLabel$ such that $\instanceVocabulary\in \petVerbalizer(\instanceLabel)$ contribute to the gradient.
In ADAPET's decoupled label loss, tokens not belonging to the subspace defined by the verbalizer are also considered in the loss calculations.
This is accomplished by computing a softmax probability normalized against the entire vocabulary.
Then, the probability of the correct tokens are maximized and the probability of the incorrect tokens (i.e., those tokens from $\petVerbalizer_{\instanceLabel'\in\setLabel, \instanceLabel'\neq \instanceLabel}$) is minimized.

The label-conditioned MLM loss of ADAPET involves further MLM training.
Specifically, given an example $\petEx$ whose label is $\instanceLabel_\petEx$, tokens from $\petEx$ are masked to give $\petEx'$.
For each label $\instanceLabel$, the model is trained to predict the correct tokens from $\petEx'$ if $\instanceLabel=\instanceLabel_\petEx$, and it is trained to \textit{not} predict the correct tokens from $\petEx'$ if $\instanceLabel\neq\instanceLabel_\petEx$.
The decoupled label loss and the label-conditioned MLM loss are summed together to form the loss for ADAPET.

A key difference between \citet{schick2021exploiting}'s PET algorithm and \citet{tam2021adapet}'s ADAPET algorithm is that the former assumes access to a large set of unlabeled text data, whereas the latter assumes access to a large set of labeled text data as a development set.
\citet{perez2021true} investigate the performance of ADAPET without access to a labeled development set, and show that its performance is worse than that of PET.
To ensure a true few-shot learning setting, we assume no access to a development set and, accordingly, choose to use the PET algorithm as, in this context, it is the state-of-the-art.

\citet{mahabadi2022perfect} develop a Prompt-Free and Efficient paRadigm for FEw-shot Cloze-based fine-Tuning (PERFECT), an alternative to PET which does not require hand-crafting patterns and verbalizers.
The PERFECT algorithm introduces task-specific adapters, which are trainable additional components added to the model architecture.
Specifically, the underlying PLM has its weights frozen, as training an entire model can be unstable and sample-inefficient when access to data is limited.
\citet{mahabadi2022perfect} also freeze the embedding layer and add a separate embedding for labels, where they fix the number of tokens allotted to each label.
Mask tokens are appended to the input sentence and the model is trained to predict the label in these mask positions.
In particular, the model is trained to predict the correct label by optimizing the label embeddings: notably, given the fixed number of tokens per label, a classifier is trained for each token and multi-class hinge loss is used over the mask positions.
As with all few-shot learning approaches, evaluating model performance or handling random variations is important for developing a good model.
For PET, this is achieved through the use of unlabeled data.
For PERFECT, a validation set is used to train a model.
However, unlike for ADAPET, \citet{mahabadi2022perfect} use half of the limited training data as their validation set: i.e., given 32 labeled instances, they set aside 16 for training and 16 for validation.
PERFECT is shown to perform worse than PET on the RAFT leaderboard~\citep{alex2021raft}. Accordingly, we do not experiment with it.

\section{Stack Overflow data queries}\label{sec:stack-overflow-queries}

For several tasks, we query for data from open-source databases.
For data scraped from Stack Overflow, the queries are SQL-based.
We query the Stack Exchange Data Explorer.\footnote{\url{https://data.stackexchange.com/stackoverflow/queries}}
As these queries are non-trivial, we provide them below in lieu of describing them in detail.

To generate pairings labeled \textit{Neutral}, we query for open posts that have been posed no later than December 31st, 2021.
We only consider open questions that have been answered at least once, as an answer ensures that the question received adequate
attention and therefore is more likely to have been closed as a duplicate if it was one.
We consider only questions tagged \textit{python}, as similar questions using different languages are not generally considered duplicates.

\begin{lstfloat}[htb]
\centering
\caption{Query for \textit{Neutral} questions}
\setstretch{1}
\begin{lstlisting}[
           language=SQL,
           showspaces=false,
           basicstyle=\ttfamily,
           numbers=none,
           numberstyle=\tiny,
           frame=None,
           commentstyle=\color{gray},
        ]
SELECT
      TOP 10000 p.id,
      p.title,
      p.creationdate,
      p.body
FROM
    Posts AS p
WHERE
      P.closeddate IS NULL
      /* use old posts (eg., before 2022)
       * to ensure enough time for 
       * duplicates to be closed */
      AND p.creationdate < '20220101' 
      /* only select questions that
       * have been answered (more likely
       * to be understood), but you might
       * want to drop this */
      AND p.answercount > 0 
      AND p.tags LIKE '%python%'
ORDER BY
    p.creationDate DESC;
\end{lstlisting}
\end{lstfloat}
As for \textit{Neutral} questions, we consider \textit{Duplicate} questions tagged \textit{python}.
\begin{lstfloat}[htb]
\setstretch{1}
\begin{lstlisting}[
           language=SQL,
           showspaces=false,
           basicstyle=\ttfamily,
           numbers=none,
           numberstyle=\tiny,
           frame=None,
           commentstyle=\color{gray},
        ]
SELECT
    TOP 10000 p.id,
    p.title,
    p.creationdate,
    p.tags,
    p.body,
    p.closeddate,
    c.name AS CloseReason,
    d.id AS dupid,
    d.creationdate AS dupcreationdate,
    d.title AS duptitle,
    d.body AS dupbody
FROM
    Posts AS p
    JOIN PostLinks AS pl ON P.id = pl.postid
    JOIN Posts AS d -- for duplicates
    ON pl.relatedpostid = d.id
    JOIN PostHistory AS h ON p.id = h.postid
    JOIN CloseReasonTypes AS C ON h.comment = c.id
    /* if postHistoryTypeId = 10 (see below)
     * then h.comment points to the ID of 
     * the close reason */
WHERE
    p.PostTypeId = 1 -- is a question
    AND pl.linktypeid = 3 -- is a duplicate
    AND p.deletiondate IS NULL -- was never deleted
    AND h.PostHistoryTypeId = 10
    AND p.tags LIKE '%python%' -- tagged "python"
ORDER BY
    p.creationDate DESC;
\end{lstlisting}
\caption{Query for \textit{Duplicate} questions}
\end{lstfloat}

\section{Replication study with other datasets}

\citet{tunstall2022efficient} report the performance of SetFit with an MPNet base versus fine-tuning (among other algorithms) in their work by experimenting with publicly available data sets.
As our results found that SetFit often underperformed or matched the performance of fine-tuning, we attempt to replicate the performance improvement of SetFit observed by \citet{tunstall2022efficient}.

\subsection{Datasets}

We consider three publicly available datasets from \citep{tunstall2022efficient}:
SST-5,\footnote{\url{https://huggingface.co/datasets/SetFit/sst5}}
AmazonCF,\footnote{\url{https://huggingface.co/datasets/SetFit/amazon\_counterfactual\_en}}
and Emotion.\footnote{\url{https://huggingface.co/datasets/dair-ai/emotion}}
For each dataset, two training set sizes are considered, constructed using a balanced sampling of 8 shots/label and 64 shots/label, respectively.
We evaluate all models on the publicly available test split of each dataset.
To ensure that the results are not skewed by an unfavourable selection of training data, for each dataset size we sample training data three times, training a model on each sample, and report the average and standard deviation of the performance.

\subsection{Experimental Setup}

As in \citet{tunstall2022efficient}, we consider SetFit with an MPNet base, denoted \setfitMpnet{}, and fine-tuning with \robertaLarge{}.
We consider two realizations of the fine-tuning procedure.
Firstly, we consider the fixed-hyperparameter fine-tuning considered throughout this work, in which the checkpoint is fine-tuned for 1,000 training steps with a batch size of 16.
We denote this as \fineTune{}.
Then, we consider the procedure used in \citep{tunstall2022efficient}.
In this case, the few-shot training data is split into an 80--20 train-validation split.
The number of training epochs is chosen inclusively between 25 and 75 as the best performer on the validation split, averaged over ten runs (to account for the impact of random starts).
This fine-tuning procedure uses a batch size of 4.
We denote this as \fineTuneSetFit{}.

It is worth noting that there are several variations of the different models in the literature for SBERT and SetFit in general.
\citet{tunstall2022efficient} use a variation\footnote{\url{https://huggingface.co/sentence-transformers/all-roberta-large-v1}} of \robertaLarge{} which is pre-trained further on sentence pairs to achieve improved performance.
In our experiments, we found that this model did not provide uniform improvement over the traditional \robertaLarge{} model.
As such further pre-trained models do not exist for \bertLarge{} nor \debertaLarge{}, we opted to keep a consistent model architecture by simply using mean pooling to construct the sentence transformer models used for our experiments.

\subsection{Numerical Results}

Table~\ref{tab:setfit-replication-results} shows the performance of each model on the considered datasets.
For 8 shots per label, we observe that both the \fineTuneSetFit{} and the \setfitMpnet{} values are all within error of the results observed in \citep{tunstall2022efficient}.
More importantly, we observe that SetFit offers a substantial performance boost in this context, including over \fineTune{} (i.e., the fine-tuning approach employed in our work).
For 64 shots per label, we observe that our results are within error for both the Emotion and the SST-5 datasets, where in each case \setfitMpnet{} outperforms \fineTuneSetFit{}. We do observe that the AmazonCF results for $\|N\|=64$ differ slightly from those in \citep{tunstall2022efficient}.
However, as we are using different test sets and averaging over three different training data samples, it is not necessarily the case that our results will match exactly.
Aside from this one outlier, we have largely reproduced the results observed in \citep{tunstall2022efficient}, notably its performance improvement over fine-tuning in a few-shot setting.




\begin{table}[htb]
\centering
\caption{Replication of SetFit versus Fine-tuning results.}
\label{tab:setfit-replication-results}
\renewcommand{\arraystretch}{1.25}
\begin{tabular}{lccc}
\toprule
 &             AmazonCF &       Emotion &          SST-5 \\
\midrule
\multicolumn{4}{c}{$\|N\|=8^*$}\\
\fineTuneSetFit{} &   $12.6\pm4.5$ &  $27.8\pm4.5$ &  $33.8\pm0.8$ \\
\fineTune{}       &   $24.0\pm7.6$ &  $40.6\pm3.3$ &  $33.6\pm2.1$ \\
\setfitMpnet{}    &  $\mathbf{45.6\pm17.1}$ &  $\mathbf{49.8\pm5.1}$ &  $\mathbf{44.4\pm0.8}$ \\
\midrule
\multicolumn{4}{c}{$\|N\|=64^*$}\\
\fineTuneSetFit{} &   $\mathbf{70.2\pm3.5}$ &  $71.2\pm3.5$ &  $50.0\pm1.3$ \\
\fineTune{}       &   $63.3\pm5.9$ &  $71.2\pm3.8$ &  $49.7\pm1.2$ \\
\setfitMpnet{}    &   $65.8\pm2.5$ &  $\mathbf{74.5\pm2.9}$ &  $\mathbf{52.4\pm1.4}$ \\
\bottomrule
\end{tabular}
\begin{tablenotes}
    \footnotesize
    \centering
    \item * $\|N\|$ indicates the number of shots per label.
    \item Reported values are MCC for AmazonCF, Accuracy for Emotion and SST-5
    \item \textbf{Bold:}~Best score for a task.
\end{tablenotes}
\end{table}

\end{document}